\documentclass[fleqn,10pt]{wlscirep}

\usepackage{braket}
\usepackage{hyperref}
\usepackage{graphicx,bm,amsmath,color}
\usepackage{hyperref}
\usepackage[flushleft]{threeparttable}
\usepackage{subcaption}
\usepackage[gen]{eurosym}
\usepackage{amsthm}
\usepackage{float}
\usepackage{placeins}
\usepackage{enumitem}  
\usepackage[english]{babel}
\usepackage{url}
\usepackage{multicol}
\usepackage{cite}
\usepackage{todonotes}
\usepackage{lineno}

\usepackage{xr}
\usepackage{soul}

\title{The emergence of segregation: from observable markers to group specific norms}

\author[1,2]{Juan Ozaita}
\author[3,4]{Andrea Baronchelli}
\author[1,2,5,6]{Angel S\'anchez}

\affil[1]{Grupo Interdisciplinar de Sistemas Complejos, Departamento de
Matem\'aticas, Universidad Carlos III de Madrid, 28911 Legan\'es, Madrid,
Spain}
\affil[2]{Unidad Mixta Interdisciplinar de Comportamiento y Complejidad Social
(UMICCS) UC3M-UV-UZ, 28911 Legan\'es, Madrid, Spain}
\affil[3]{City, University of London, Department of Mathematics, London EC1V 0HB, UK}
\affil[4]{The Alan Turing Institute, British Library, 96 Euston Road, London NW12DB, UK}
\affil[5]{Institute UC3M-Santander for Big Data (IBiDat), Universidad Carlos III de Madrid, 28903 Getafe, Madrid, Spain}
\affil[6]{Instituto de Biocomputaci\'on y F\'\i sica de Sistemas Complejos (BIFI), Universidad de Zaragoza, 50018 Zaragoza, Spain}

\begin{abstract}
Observable social traits determine how we interact in society and remain pervasive even in our globalized world. While a popular hypothesis states that they may help promote cooperation, the alternative explanation that they facilitate coordination has gained ground in recent years. Here we explore this framework and present a model that investigates the role of ethnic markers in coordination games. We consider fixed markers characterizing agents that use reinforcement learning to update their strategies in the game. For a wide range of parameters, we observe the emergence of a collective equilibrium in which markers play an assorting role. However, if individuals are too conformists or greedy, markers fail to shape social interactions. 
These results extend and complement previous work focused on agent imitation and show that reinforcement learning is a good candidate to explain many instances of ethnic markers. 
\end{abstract}

\begin{document}

\flushbottom
\maketitle
\thispagestyle{empty}

\section*{Introduction}\label{introduction}

In spite of the globalized nature of our societies, and our access to knowledge and information about every other population in the world, ethnic groups behaving according to disparate social norms or conventions are ubiquitous \cite{boyd1987}. According to Barth\cite{barth1969}, people identify themselves, and are identified by others, as belonging to certain ethnic group by means of culturally transmitted features, such as 
wealth symbols, language, culture and artistic forms, dress style or cuisine. All these attributes have in common that they are external and can be seen,  evaluated and acted upon by any other member of the population. In turn, mechanisms based on social categorization and parochialism work to maintain them.\cite{bernhard2006}

Several researchers argued that ethnic groups are the basic locii for cooperation, enabling individuals to profit from cooperative exchanges depending on the observation of the traits displayed by others. \cite{nettle1997,vandenberghe1981,cohen2012} However, interpretations in terms of cooperative or altruistic predisposition towards similarly marked individuals has been challenged based on the possibility of free-riding. \cite{price2006} An alternative explanation solves the free-riding problem by maintaining that the social role of ethnic markers can be to facilitate coordination rather than cooperation.\cite{gil-white2001,mcelreath2003shared,moya2013} While there is contradictory experimental evidence on this hypothesis, \cite{chen2009,simpson2006,fersthman2001,koopmans2014} research specifically designed to test between these two options seem to favor the coordination interpretation.\cite{jensen2015}

In this paper, we focus on the role of ethnic markers on norm formation, modelled as a game of coordination. \cite{lewis1969convention,steels1995self,baronchelli2006sharp,centola2015spontaneous} In other words, we assume that society is regulated by norms that reward interactions betweeen individuals that share beliefs about how people should behave. \cite{skyrms2004,baronchelli2018consensus} We consider fixed markers and introduce the evolution of action choices through reinforcement learning,\cite{bush1955,macy2002learning,izquierdo2008} realistic in many instances of interaction between humans.\cite{cimini2014,cimini2015}. We do so in the framework introduced by McElreath {\em et al.}\cite{mcelreath2003shared} consisting of a unique binary marker used by individuals to choose their strategy in social interactions based on proportional imitation. This characterization of strangers based on ethnic markers is thus used to determine whether or not there are shared social norms with them. \cite{schelling1960,sugden1995}

The rationale for exploring reinforcement learning as an alternative to proportional imitation is that, while the first has been criticised for assuming excessive flux of information in the population, the latter allows to achieve a cooperative equilibrium by using only individual information. \cite{flache2002stochastic,erev2001,bendor2001} In our model, each agent is defined by an individual parameter called the aspiration. This parameter defines how an outcome makes it more likely to choose a given action, when the payoff it leads is larger than the aspiration (positive stimulus) or on the contrary makes it less likely (negative stimulus). In the following sections, we explore the role of this parameter, and the other parameters already present in Ref.\ \citen{mcelreath2003shared}.

\section*{Model}\label{Fixedaspirations}

In this section we introduce our model for the marker's mechanism incorporating reinforcement learning\cite{bush1955,macy2002learning,izquierdo2008} as our paradigm for the evolution of the agents decisions. The key feature of this dynamics is the aspiration parameter, which agents use to evaluate the outcome of a round. If the outcome is above (below) their aspirations, they generate a positive (negative) stimulus and tend to repeat (avoid) their previous action. It has been shown in Ref.\ \citen{macy2002learning} that aspirations that generate positive and negative stimula tend to promote cooperation, so we will consider different aspiration values to analyze the role of this parameter in the evolution of actions when markers mediate a coordination game. 

In our model, agents are characterized by the following parameters:

\begin{itemize}
\item \textbf{Behavior:} It is the action chosen for the coordination game. It may take two different values $\left\lbrace 0,1\right\rbrace$.
\item \textbf{Marker:} It is a visible characteristic which identify the interaction. It may take two different values $\left\lbrace 0,1\right\rbrace$. In our version markers will be immutable, modeling observable social traits that either do not change either change in a very slow timescale. 
\item \textbf{Aspiration:} The payoff expected by the agent, which will define the stimulus that it receives from the interaction.
\item \textbf{Probability vector:} The probability for an agent to choose a behavior according to the subject's marker. It is divided in normalized categories, probabilities to choose an action when interacting with an agent with the same marker, and the same for the case of an agent with the opposite marker, i.e., 
\[
p_{=,0}+p_{=,1}=1 
\]
	\[
p_{\neq,0}+p_{\neq,1}=1
\]
\end{itemize}

On the other hand, we have included in the model several variables, mostly taken from  the pionnering about markers in Ref.~\citen{mcelreath2003shared}, that characterize the population and their interactions as a whole: 

\begin{itemize}
\item \textbf{$e$:} Probability for the interaction to be marked, modelling the willingness of the agents to choose someone with the same marker. This parameter takes into account the tendency to bias interactions towards the marker we share with other members from our population, or in other words, the degree of homophily.\cite{mcpherson2001}
\item \textbf{$\delta$:} Extra payoff for succesful coordination. In the following sections, we will set $\delta=0.5$. 

\end{itemize}

The model consists of $N$ agents connected in a certain way, that behave according to the following dynamics:
\begin{enumerate}
\item \textbf{Select type of interaction:} With probability $e$, the individual interacts with another one chosen at random without making any reference to the marker, and with probability $1-e$ the individual interacts with another one whose shares her same marker.  This implies that interactions with others sharing one's marker take place with probability $1-e/2$. Once the type of interaction is selected, a random couple that fulfill the chosen criteria. 
\item \textbf{Select individual actions and play the game:} Both interacting agents select a behaviour according to their probability vector and play the coordination game accordingly.
\item \textbf{Collect payoff and update probability vector:} Both agents collect their payoff. If this payoff satisfies their expectations, they generate a positive stimulus, and vice-versa. A positive stimulus encourages the agent to repeat the same action and a negative one repel him. These dynamics are captured in the following equations, extracted from \cite{macy2002learning}:
\begin{equation}\label{eq6}
p_{a,t+1}= \left\{ \begin{array}{lcc}
             p_{a,t}+(1-p_{a,t})ls_{a,t} &   if  & s_{a,t}\geq 0 \\
             \\ p_{a,t}+p_{a,t}ls_{a,t} &  if & s_{a,t}<0 \\
             \end{array}
   \right.
\end{equation}
Here, $l$ is the learning rate\cite{macy2002learning}, that controls the speed of adaptation of the agents. 
Subsequently, the probability vector is normalized for each category, completing a timestep.
\end{enumerate}

\section*{Results}

We present here our main results for the model above. First of all, we will consider the same value for all the agents aspirations and we will characterize the different phases of it. Then, we will study variations on the parameters exposed before. We will also analyze the effect of modifying the payoff matrix (imperfect coordination), learning rate variations ($l$), assortment effects (more than one initial option for the aspiration), migration effects and variations on the bias of the interaction ($e$). Unless otherwise specified, the values of the main parameters and variables of the model are as follows:
Learning rate $l=0,5$ (fast learning), $e=0.5$ (interaction is biased towards individuals with the same marker, and a unique population of $N=500$ agents. We run simulations for 4\,000 interactions per agent (in total, $3\cdot 10^{6}$), and statistical averages and plots have been made with 100 simulations with markers and behaviors chosen at random.

\subsection*{Baseline}

\begin{figure}[htp]
  \includegraphics[width=.23\linewidth]{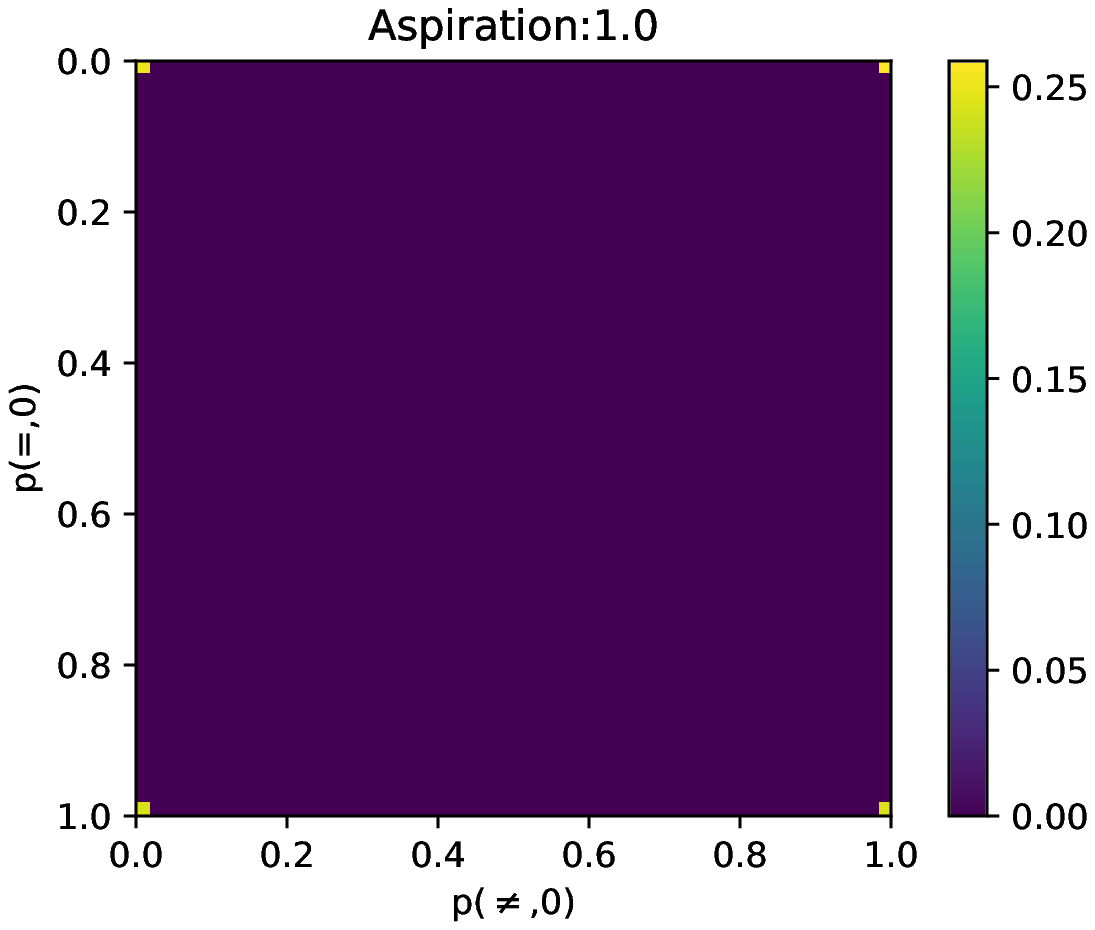}  
  \includegraphics[width=.23\linewidth]{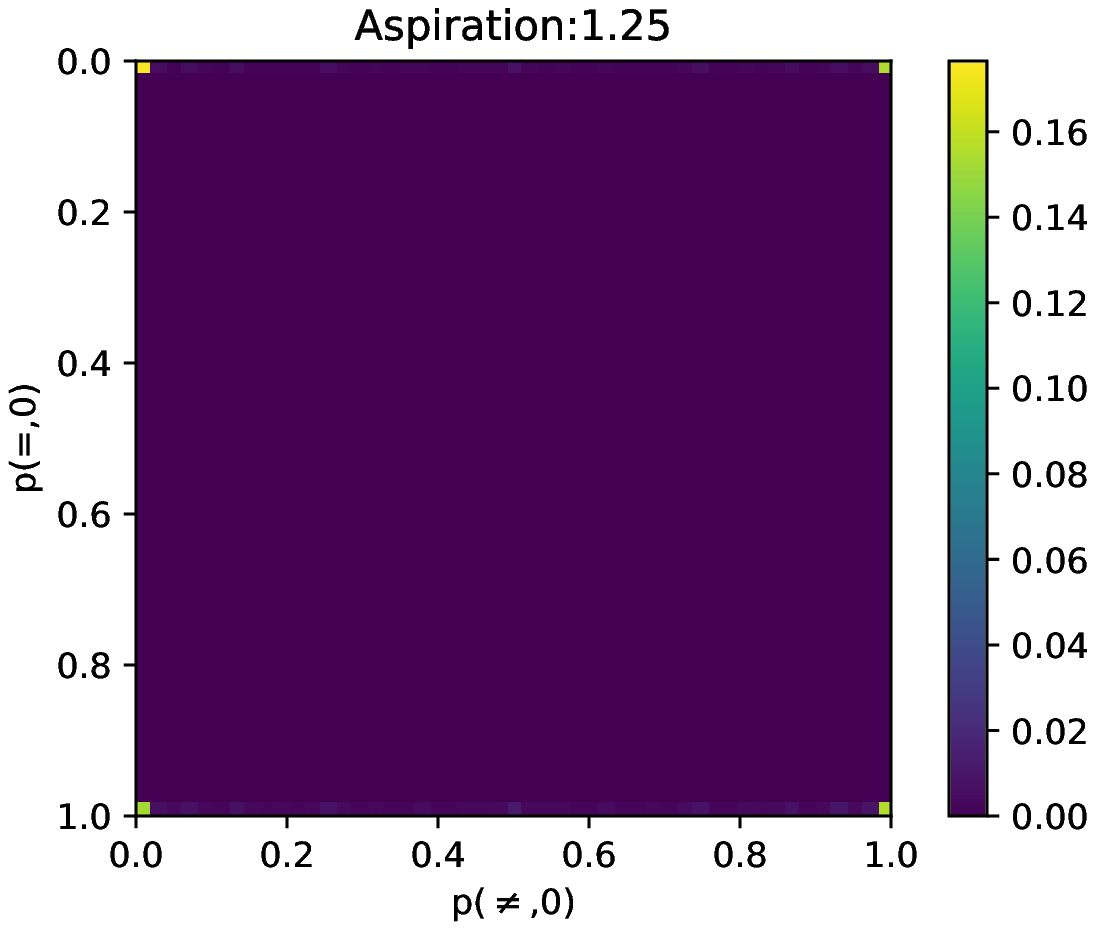}  
  \includegraphics[width=.23\linewidth]{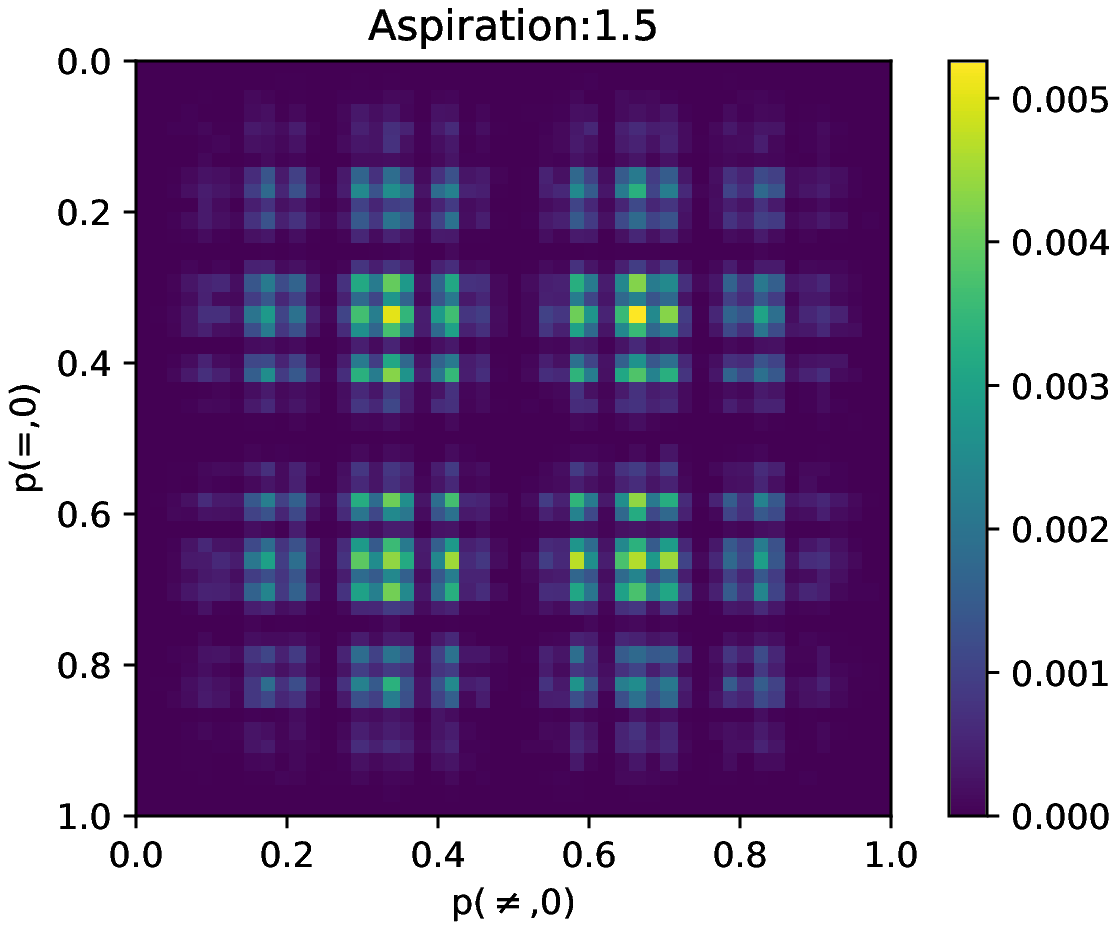}  
  \includegraphics[width=.23\linewidth]{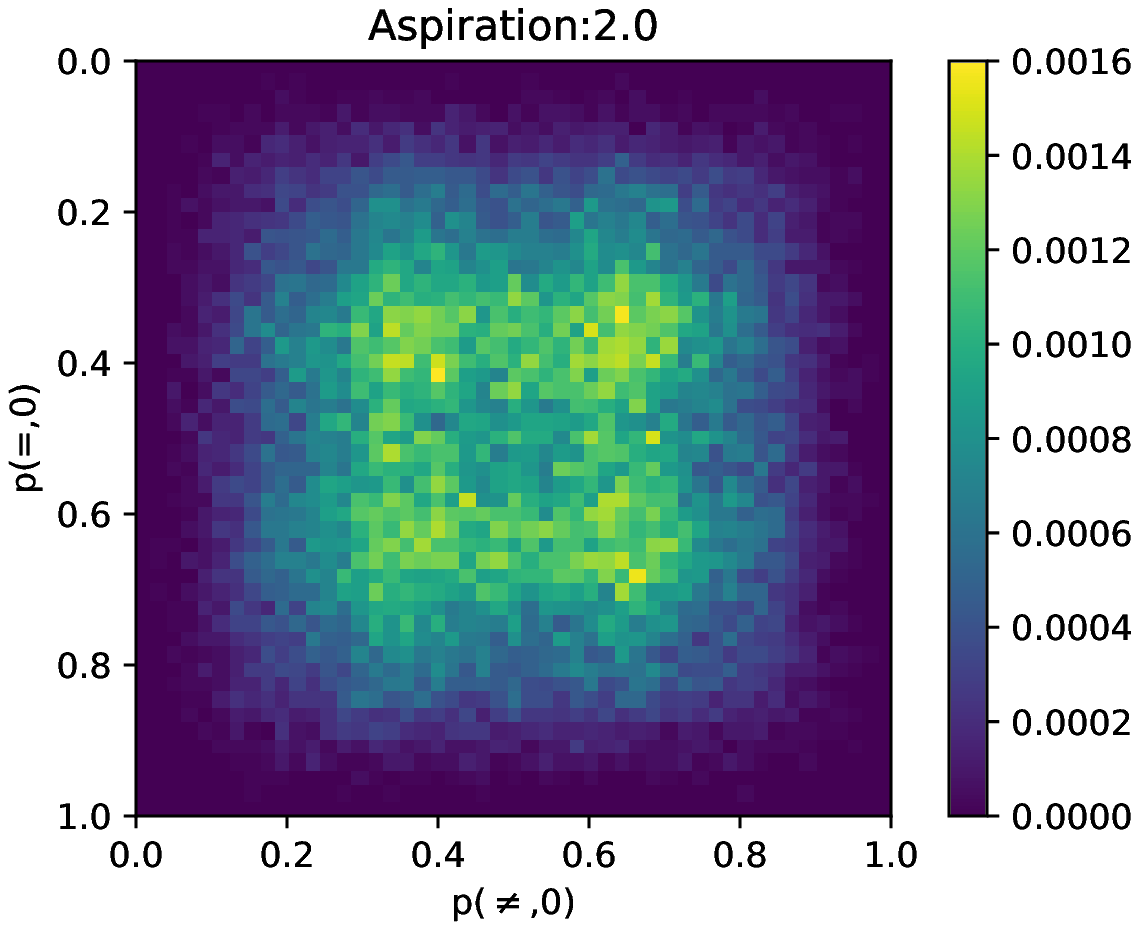}  
  \caption{2D histograms for the probability vectors of the population.  Shown is the fraction of agents with specific values of choosing action 0 when facing another agent with different marker (horizontal axis) and when facing another agent with the same agent (vertical axis). From top to bottom and from left to right, aspiration levels are  $1$, $1+\delta/2$, $1+\delta$, $1+2\delta$.}
\label{fig:Fig1}
\end{figure}

 We start by looking at the analysis of the probability vector phase space for different values of the aspirations. Specifically, we consider the cases $A_{i}=\left\lbrace 1, 1+\delta/2, 1+\delta, 1+2\delta \right\rbrace$, which represent aspiration levels at the minimum payoff(and therefore always satisfied or neutral), an intermediate value between payoffs, the maximum payoff, and a level above all payoffs (which leads to negative stimulus for all outcomes), respectively. Results are shown in Fig.\ \ref{fig:Fig1}.

\begin{figure}[htp]
  \centering
  \includegraphics[width=.5\linewidth]{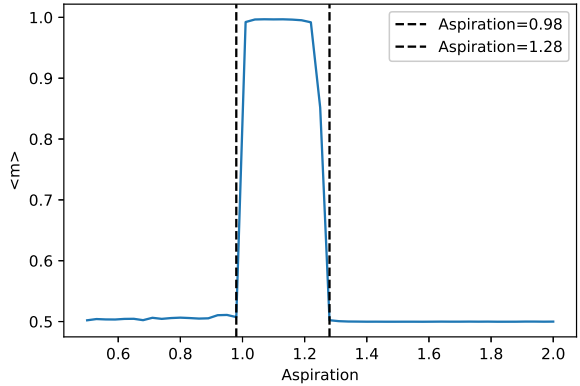}  
\caption{Coordination ratio $m_{i}$ for different values of the aspiration, averaged over agents, time and simulations.}
\label{fig:Fig2x}
\end{figure}

We can distinguish two main kinds of behavior depending on the value of  aspiration: 
For $A_{i}\leq 1+\delta/2$ (this limit will be prooved in Fig. \ref{fig:Fig2x}), the system is dominated by positive stimula. Reinforcement learning leads agents to deterministic behavior, sticking to one of the actions for each of the two possible interactions, individuals with the same marker and with different markers. On the contrary, when
$A_{i}>1+\delta/2$, the learning process is dominated by negative stimula, and agents behave more randomly, meaning for both categories agents may choose one or other strategy with nonzero probability. We will see that this behavior appears at $A_{i} = 1+\delta/2$ and can be seen explicitely in the last subfigure. 

In order to shed further light into the different regimes, we introduce the \textit{ratio of coordination} $m_{i}$ defined as
\begin{equation}\label{eq7}
m_{i}=\dfrac{\mbox{\rm no.\ of coordinations}}{\mbox{\rm no.\ of interactions}}
\end{equation}

The ratio of coordination allows us to monitor how succesful the average player is in coordinating with every individual she meets. In Fig.\ \ref{fig:Fig2x} we show the values of this parameter averaged over the length of the run and over simulations for an interval of values for the aspiration level. 
We identify three regimes. 
The first regime arises when $A_{i}\leq 1$: All stimula are positive (or neutral, in the extreme case). This means that every action  encourages the subject to repeat the same action. If we let this dynamics evolve, the subjects end up organized in the four possible combinations of $\left\lbrace p_{=,0},p_{\neq,0} \right\rbrace$. As the agents always receive the same stimula, resulting in no correlation strategy-marker. We will refer to this type of behavior as {\em frequentists} referring to the fact that their choices depend on the agent and its sequence of interactions. As a consequence, the ratio of coordination is around 0.5, as expected in such a random setting. 

The second regime arises when we increase the aspiration further, $1<A_{i}<1+\delta/2$. Here, both positive and negative stimula exist but the formare dominate. Agents are again clustered, but this time $p_{=}$ is marker related and $p_{\neq}$ is unique for all the population. If both behaviors (strategies) are the same, homogeneity is promnoted. At the end of the simulations, we observe that agents develop intra-marker correlations (collective organized strategy within each marker subgroup) and inter-marker correlations (collective organized strategy for interactions between marked subgroups). Thus, agents have a criteria for both of the categories. We will call them {\em learning} agents, as they do learn to coordinate; note that the ratio of coordination is almost 1, the difference being due to mistakes or, equivalently, fluctuations during the learning process. 

Finally, the third regime occurs when  
$1+\delta/2\geq A_{i}\leq 1+\delta$. We enter the stochastic behavior: Positive and negative stimula exist, but negative dominate. Probability vectors are uniformly distributed in the phase space. Agents do not share a criterium nor have a clear strategy for any of the categories. We will call them {\em random walkers}. The ratio of coordination is again 0.5, as in the case of frequentists. 

\subsection*{Variations on the baseline} 

Having described in detail the phenomenology we observe on our baseline model, in this subsection we are going to study the role of the parameters we have included in this model by varying them one by one and analyzing the change in the qualitative results.

\paragraph{Variations on the perfection of the coordination game.}
What happens if coordination equilibria are not equivalent? We explore this scenario by modifying the payoff matrix as follows.
\begin{equation}\label{eq8}
\begin{pmatrix}
1+\delta+a_{1} & 1-b_{1}  \\
1 & 1+\delta 
\end{pmatrix}
\end{equation} 

\begin{figure}[htp]
\begin{subfigure}{.5\textwidth}
  \centering
  \includegraphics[width=.8\linewidth]{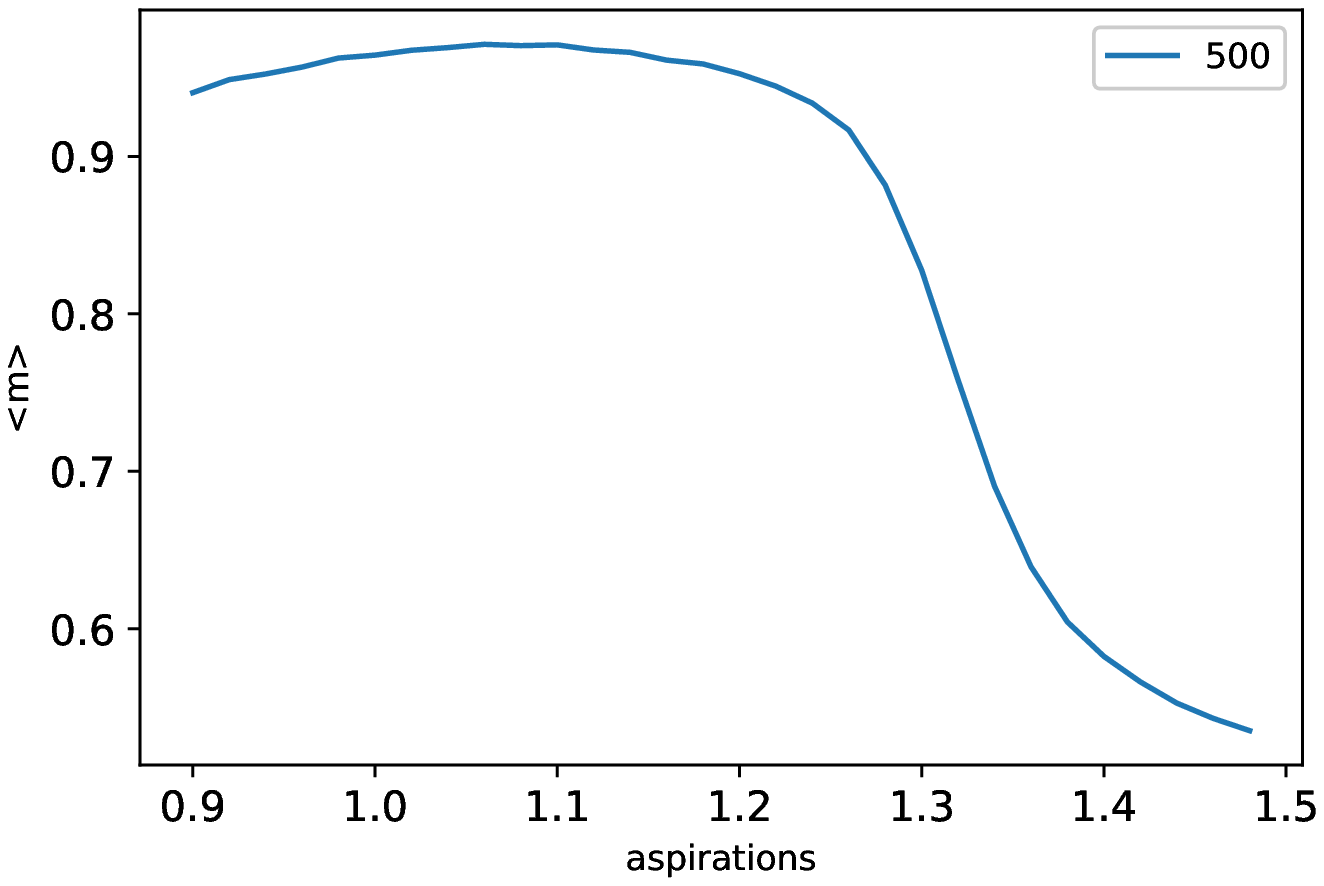}  
  \label{fig:Fig31}
\end{subfigure}
\begin{subfigure}{.5\textwidth}
  \centering
  \includegraphics[width=.8\linewidth]{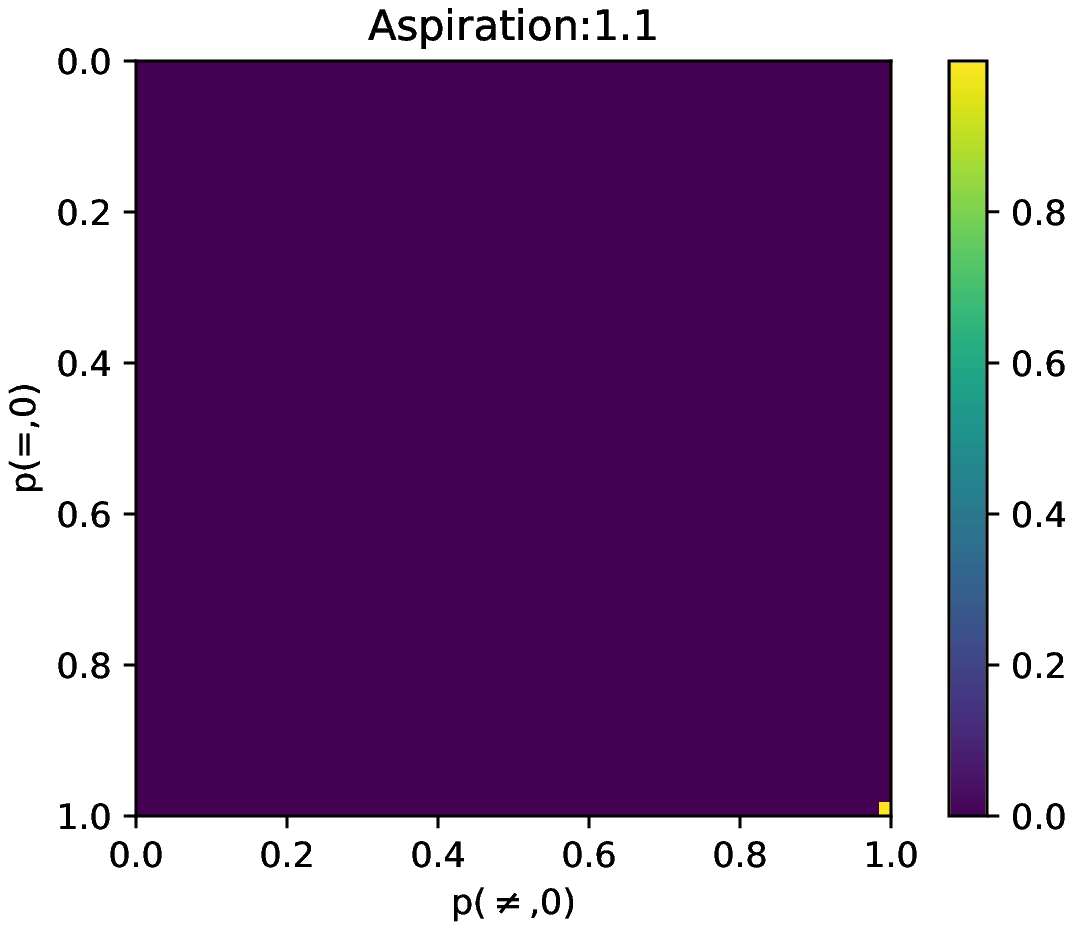}  
  \label{fig:Fig32}
\end{subfigure}
\caption{Evolution of the average ratio of coordination for the asymmetric coordination problem with $a_{1}=b_{1}=\delta/2$ (left) for different aspiration levels, and  2D histograms for the probability vectors of the population (right) for $A_i=1.1$.}
\label{fig:Fig3}
\end{figure}

We consider, as an example, $\delta=0.5, A_{i}=1+\delta/2, a_{1}=b_{1}=\delta/2$, obtaining the results shown in Fig.\ \ref{fig:Fig3}.
Agents choose the Pareto-dominant equilibrium if their aspiration level is not too large: reinforcement learning agents  organize around this strategy, maximizing their individual and global payoff. The 2D histogram shows clearly that all agents choose the best equilibrium irrespective of the markers. Therefore, markers play no role when different outcomes have different payoffs because higher payoffs provide higher stimula, which is the key element for an agent in order to arrive to the final equilibrium. Thus, markers are only relevant to self-organize the system when there are several equilibria with the same payoff, clustering agents in categories $\left\lbrace 0,=,\right\rbrace$, $\left\lbrace 0,\neq,\right\rbrace$, $\left\lbrace 1,=,\right\rbrace$,$\left\lbrace 1,\neq,\right\rbrace$, which is not possible in the absence of markers. Categories arising from markers define the new collective agreements or conventions that are present in the population. It has to be kept in mind that in addition to the two equilibria being equally benefitial, agents must have the right aspiration value, for positive stimula to lead to the evolution of coordinated behavior.

\paragraph{Variations on the learning rate ($l$).}

\begin{figure}[htp]
  \centering
  \includegraphics[width=.48\linewidth]{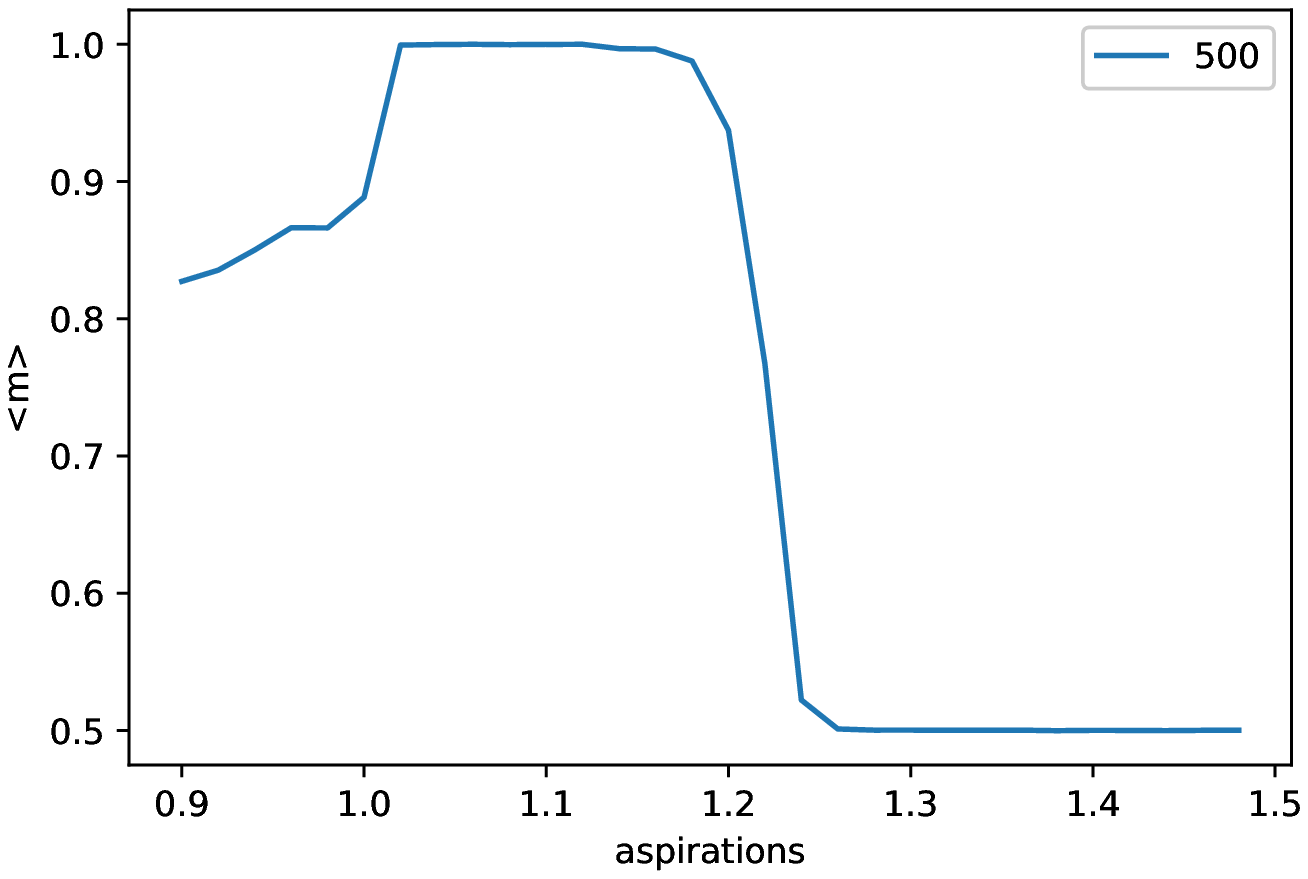}
  \includegraphics[width=.48\linewidth]{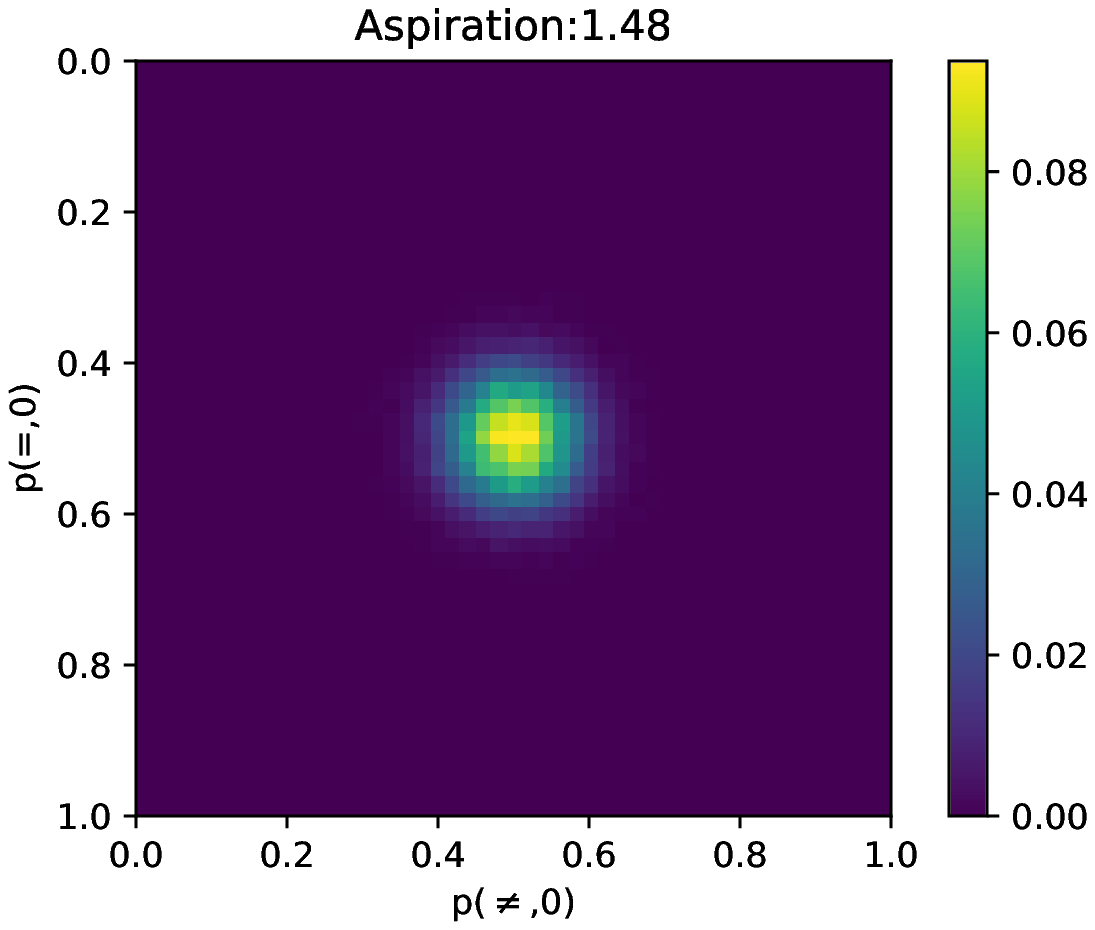}    
\caption{Left: Coordination ratio $m_{i}$ for different values of the aspiration, averaged over agents, time and simulations. Right: 2D histograms for the probability vectors of the population.  $l$=0.05.}
\label{fig:Fig4}
\end{figure}

In the baseline simulations, we have used fast learning, i.e., agents react strongly to the stimulus. Let us now look at the possibility that qualitative results, like the marker correlations, are the same if the learning rate is slower ($l=0.05$). The results are summarized in Fig.\ \ref{fig:Fig4}.
The main difference with the previous case is the higher level of coordination among the populations with low aspirations. With a low learning rate, the influence of stimuli is very weak. Agents need more interactions to achieve a fixed probability vector, which is expected for the equilibrium in agents with low aspirations. This increase in the number of interactions promote a more homogeneous equilibrium.
It is interesting to note that if we could take the limit $l\rightarrow 0$ our agents would take infinite time to arrive to an equilibrium, but it would be entirely homogeneous. We can conclude then than, at least in the variable $m_{i}$, the qualitative difference in behavior around $A_{i}=1$ can become continuous if we choose the proper learning rate.
The variation on the learning rate can be also noted in the high aspirations region. As the learning rates are smaller, random-walkers stay trapped in the center of the "phase space". 

\paragraph{Variations on the bias of the interaction ($e$).}

Another important feature of our model  is the bias on the interaction. In order to discuss the effects of this, we consider an  example of a trajectory to check the evolution of the coordination rate in the population. We choose a population with a global aspiration $A_{i}=1,1$ to promote coordination, and check, for a slow learning rate ($l=0,05$), the difference between $e=0,5$ and $e=1$.
We can see that there are two different coordination velocities, represented by the different slopes of the curve. There are two stages in the process of arriving to the equilibrium: In the first one, the system is not organized and it is building both criteria, the intra-marker one and the inter-marker. In the second one, the intra-marker criteria exists, but the agents  do not know yet what to do when they interact with someone with a different marker. 
The bias can be also seen in the phase space of probability vectors phase space. If we study the statistics of the previous case, to show the comparison. 
We can see that the biased case has some kind of transition in two stages, from full coordination to random walkers. Firstly, the inter-correlation is broken so the agents spread around one of the axis. In a second stage, the intra-correlation is destroyed and we arrive to the homogeneously distributed phase space. On the other side, if there is no bias, both correlations are destroyed at the same time.
\begin{figure}[htp]
  \includegraphics[width=.33\linewidth]{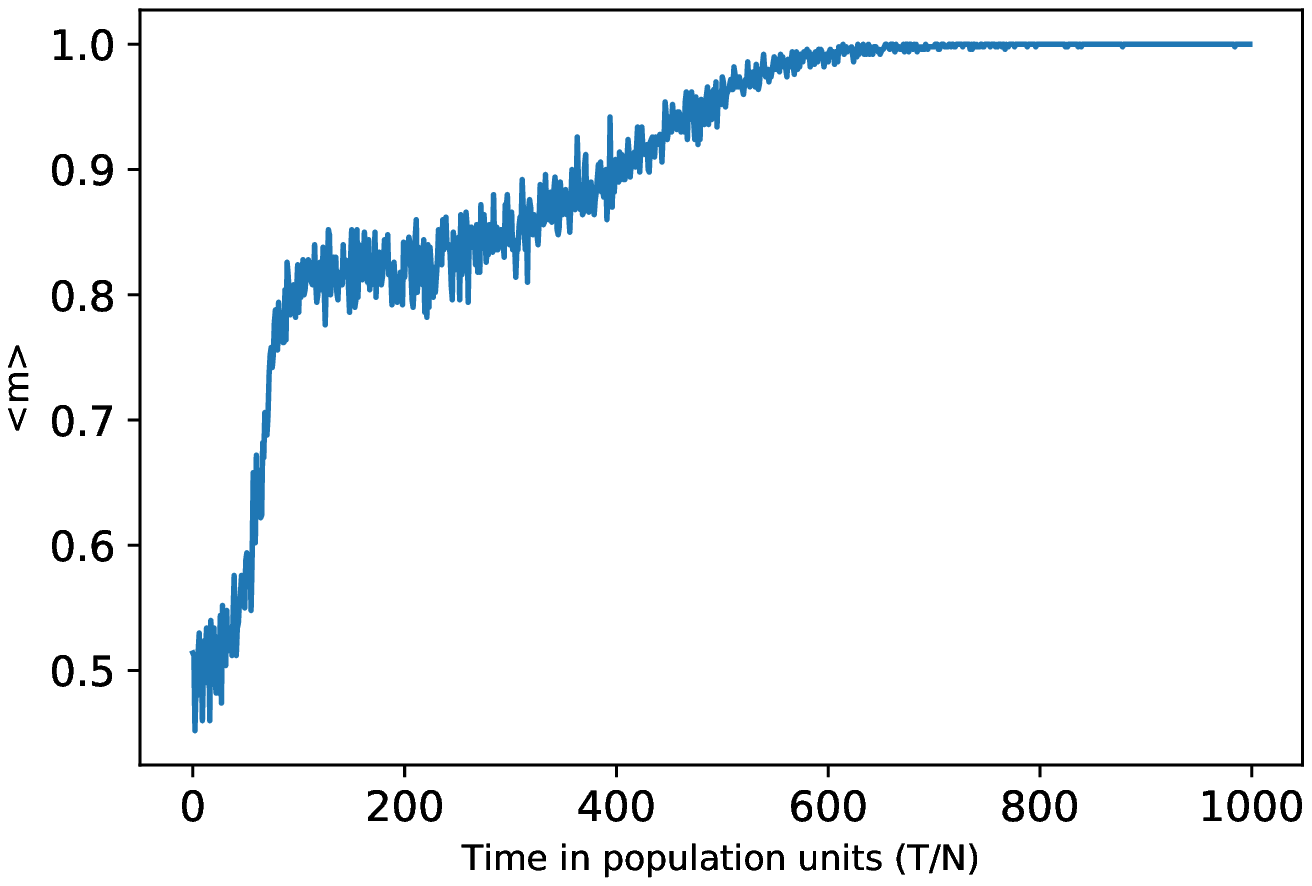}  
  \includegraphics[width=.33\linewidth]{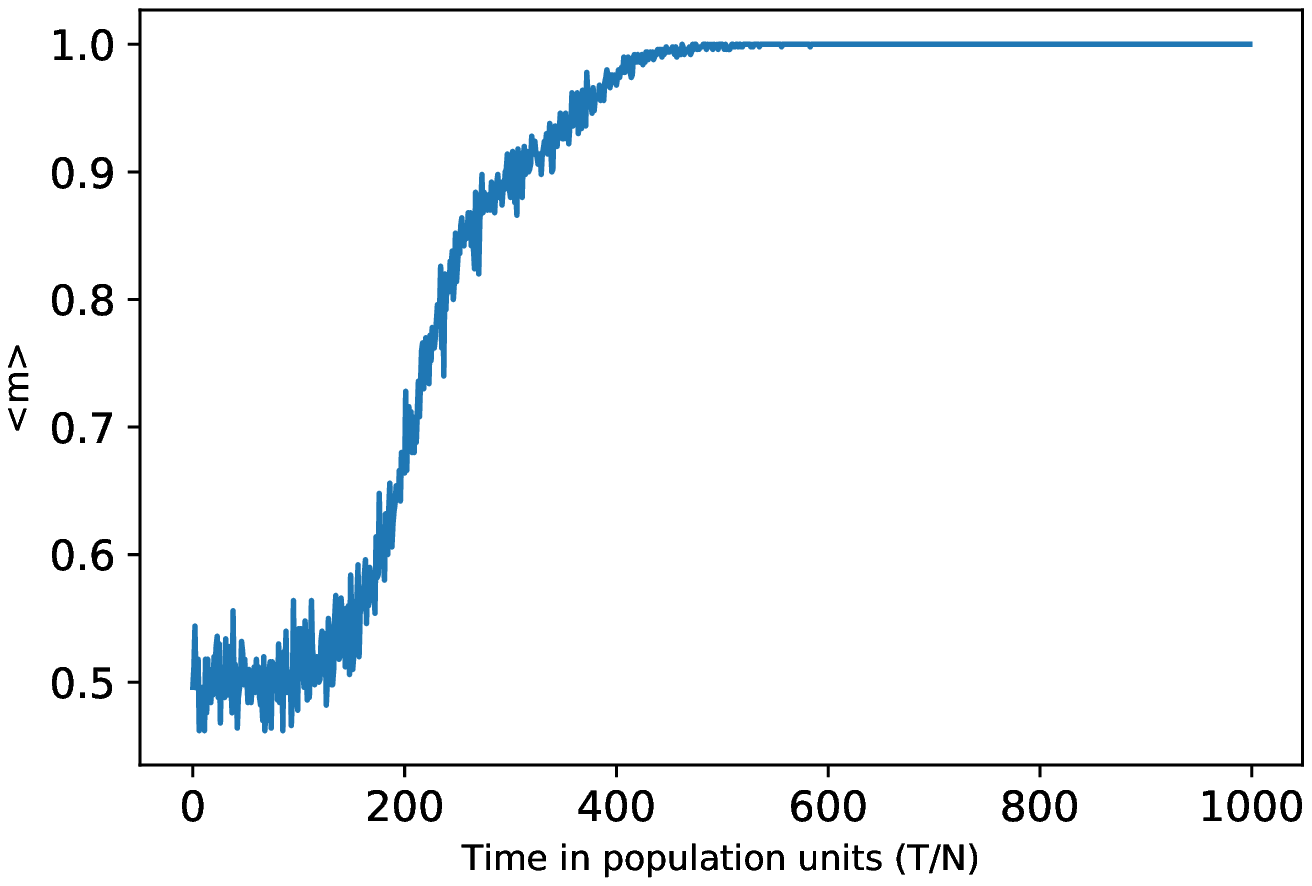}  
  \includegraphics[width=.33\linewidth]{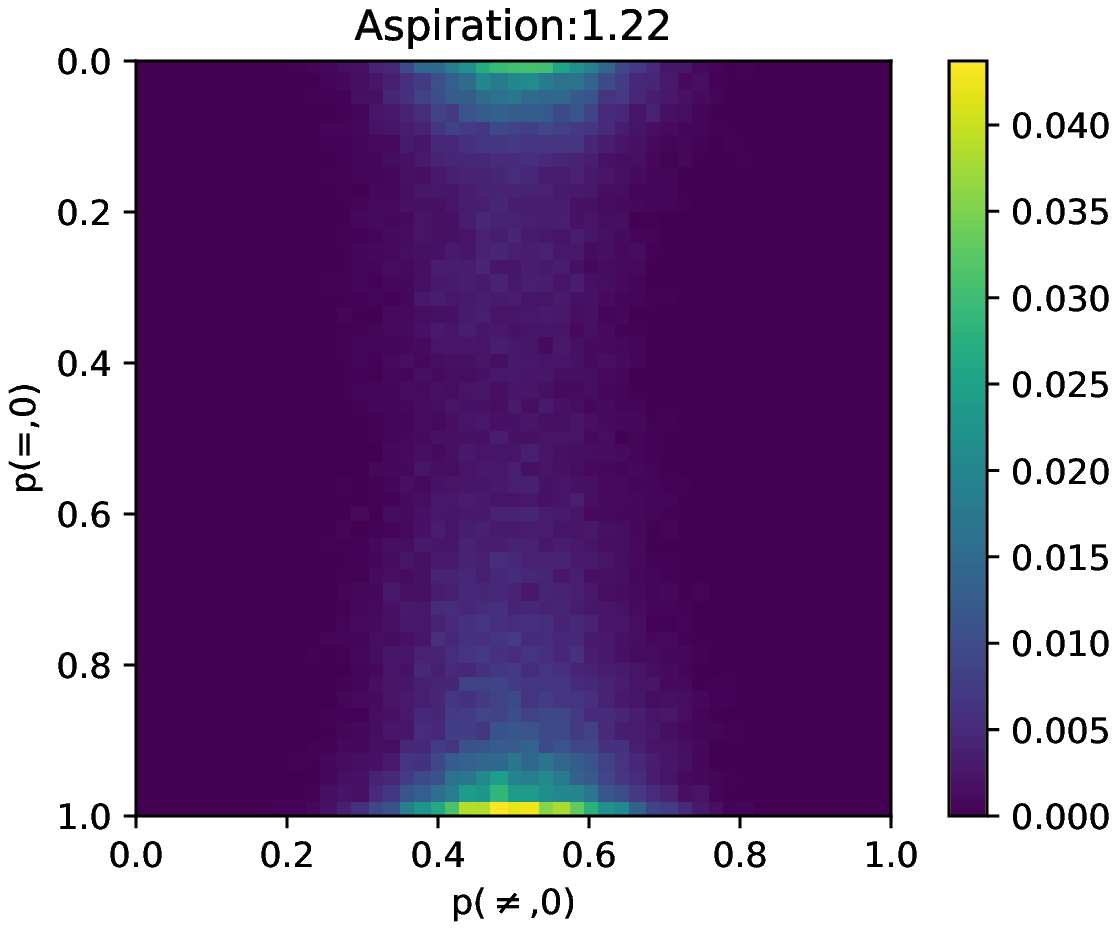}  
    \caption{Average coordination ratio for $A_{i}=1.1$ and $e=0,5$ (left) or $e=1$ (center). Right: 2D histograms for the probability vectors of the population.  $A_{i}=1,22$}
\label{fig:Fig7}
\end{figure}

Further information can be obtained from the coordination picture near the transition between learning agent and random walker. 
As it could be seen also in the probability vector phase space, in the biased case the weak correlation is destroyed and the stronger one stays. In the unbiased case, both transition persist at the edge of the transition. And, as we can see, the coordination distribution has a positive skew in the biased case and a negative one in the unbiased. We can stay then that the strength of the correlation is interaction based, as weakening one of the correlations makes a higher probability of obtaining lower levels of coordination.   
\begin{figure}[htp]
  \includegraphics[width=.48\linewidth]{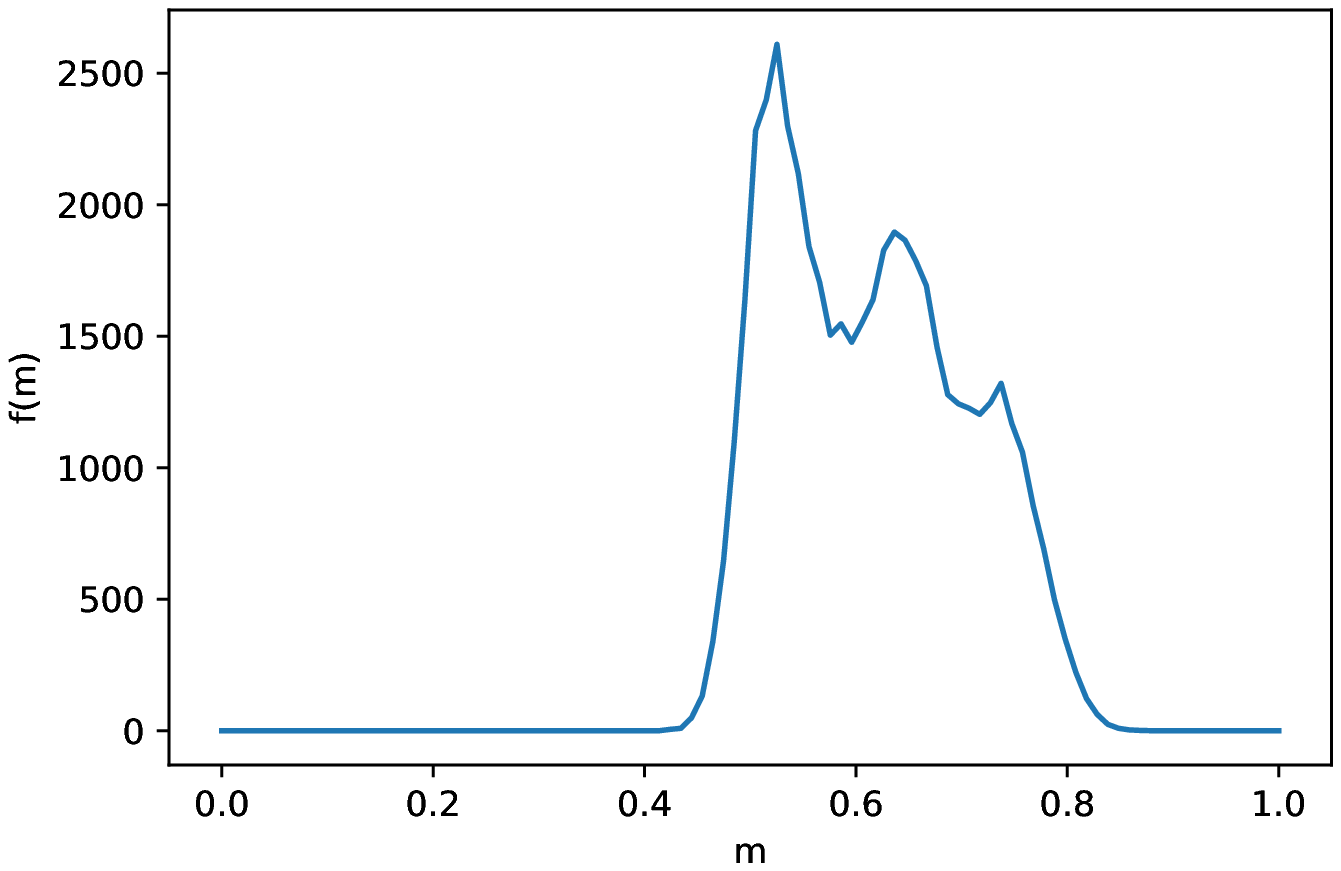}  
  \includegraphics[width=.48\linewidth]{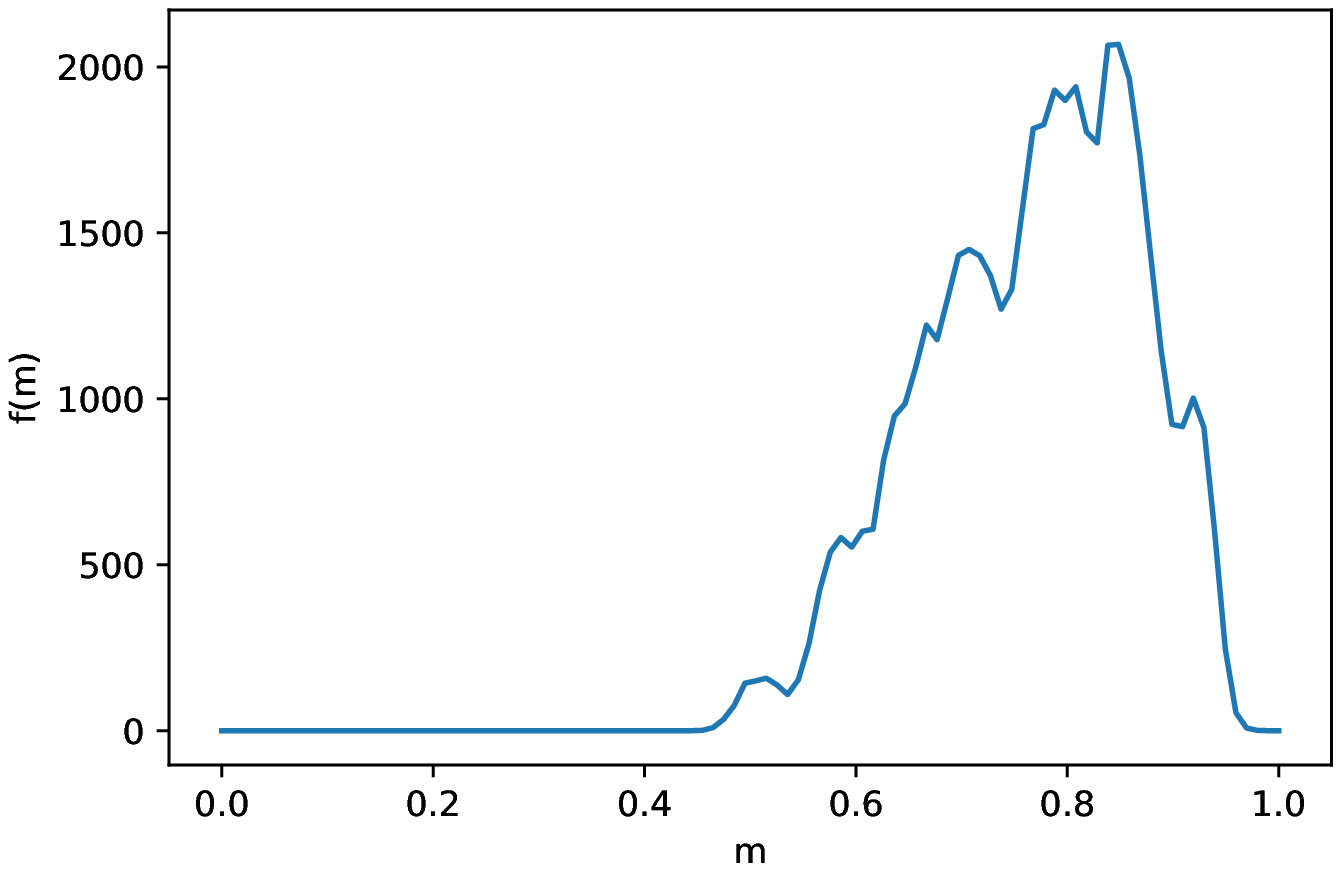}  
  \caption{Histograms of the coordination ratio for $A_{i}=1.1$ and $e=0,5$ (left) or $e=1$ (right).}
\label{fig:Fig8}
\end{figure}

\paragraph{Heterogeneous aspirations.}

In order to study fragmented populations we are going to choose three different values for the aspirations, representing the three main categories we have analyzed in the former section. These values will be $A_{i}=\left\lbrace 0.8,1.1,1.5 \right\rbrace$ and they will appear in the population with different proportions. 
We will study populations with two different categories of aspirations in different proportions ($25\%,50\%,75\%$) and populations with three different categories of aspirations in an equal proportion. 
These set ups will be studied under conditions of fast learning rate ($l=0.5$) and marked-biased interaction($e=0.5$), with all the simulation parameters as explained in the former section, as thermalization times have not changed. 

\begin{figure}[htp]
  \includegraphics[width=.48\linewidth]{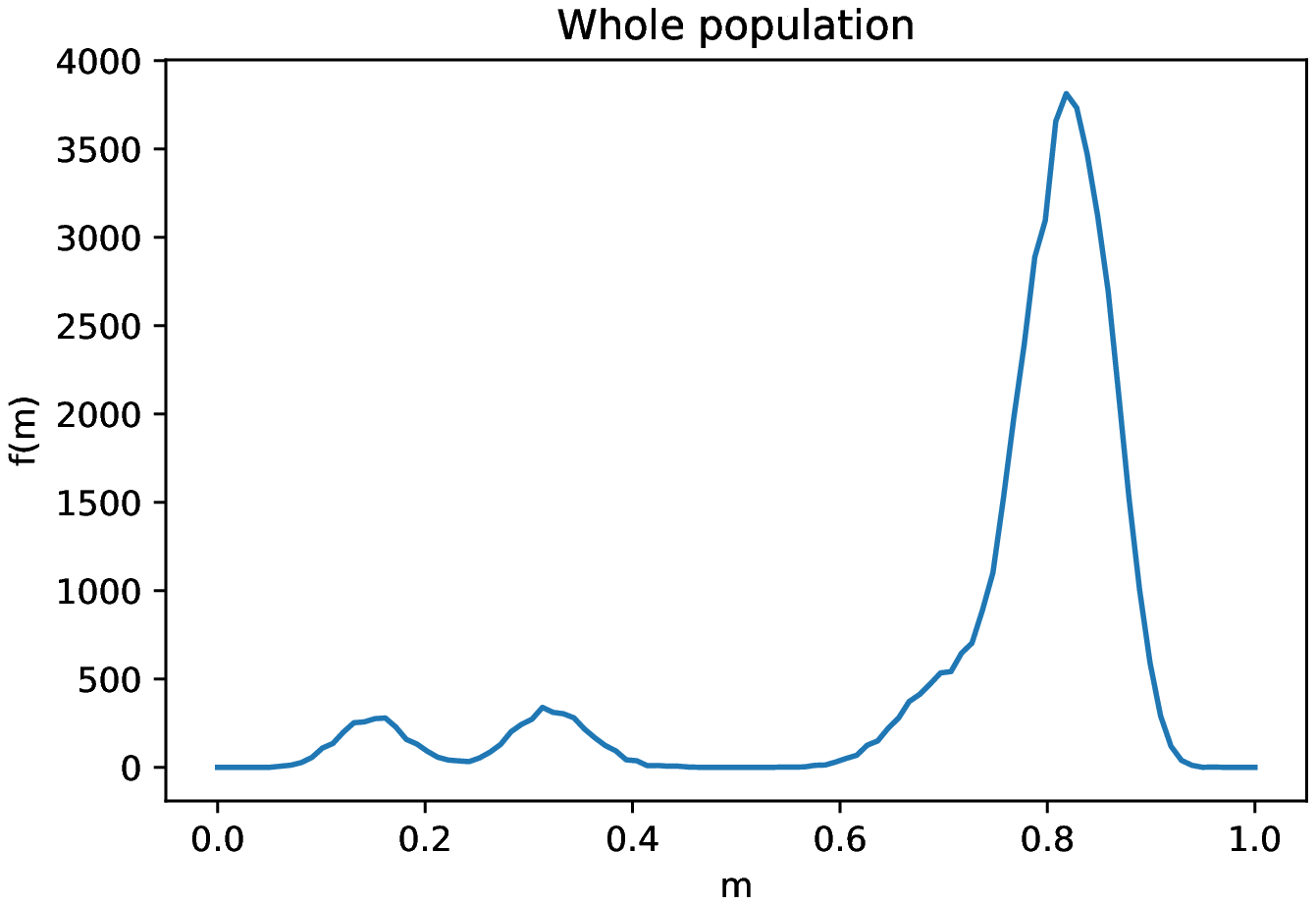}  
  \includegraphics[width=.48\linewidth]{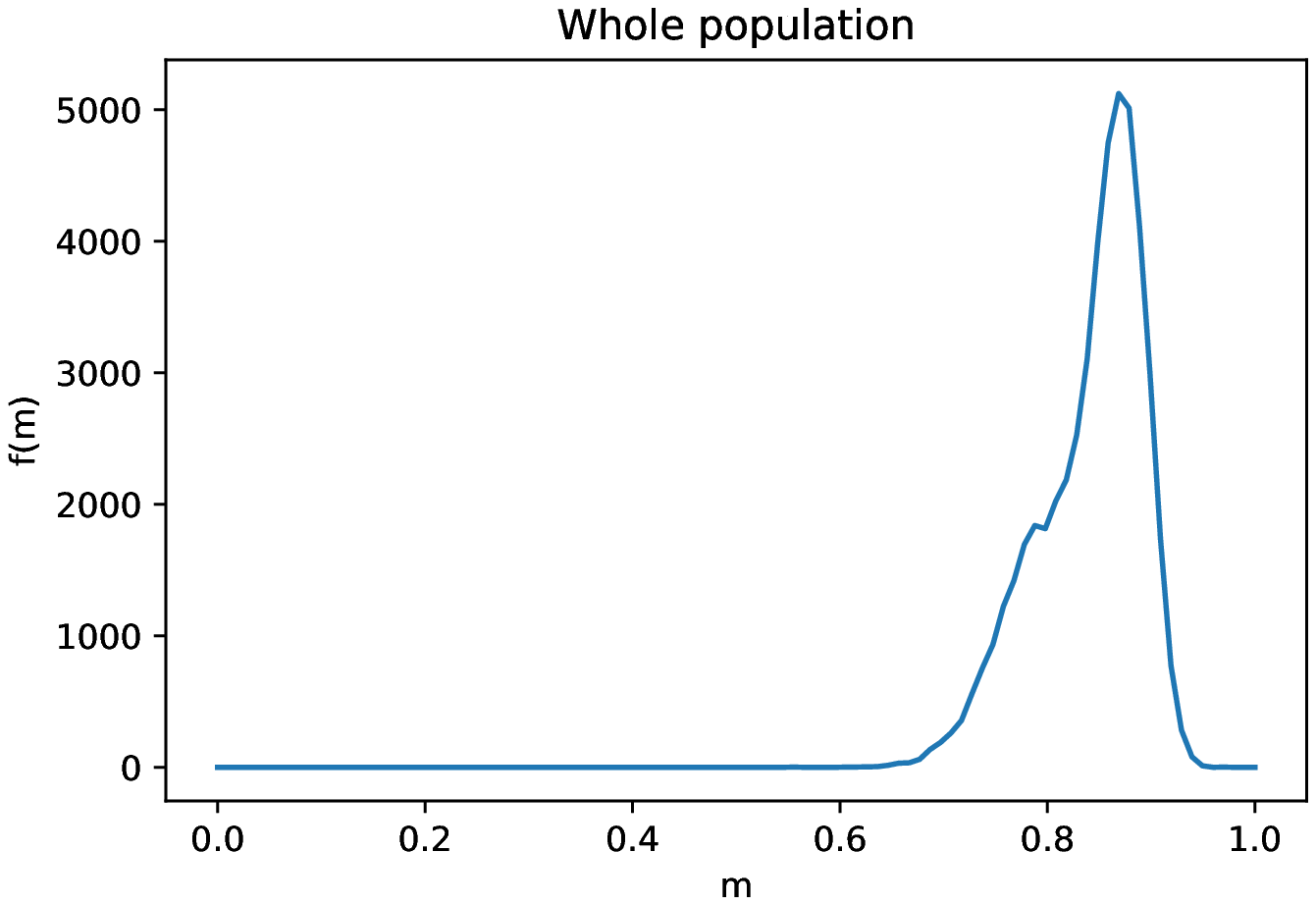}  
  \caption{Coordination ratio for a a frequentist/learning (left) or a random walker/learning agent (right) population, with the ratio (1:3).}
\label{fig:Fig9}
\end{figure}

In general, when learning agents are mixed with other type of agents they lose the full coordination they achieve when they are on their own, as described in the case of the baseline simulations. However, there are other consequences of this mixed population that change depending on the precise compostition considered. Thus, if learning agents are a majority (75\% of the population), intra and inter-marker correlation still exist but they are not perfect. This means that $0,5<p{=,i}<1$,$0,5<p{\neq,i}<1$, so their most probable option is to choose the correlated probability, but they could break the rule (partial correlation). In turn, the global level of coordination depends on who are they mixed with:  If they are mixed with frequentists, Fig.\ \ref{fig:Fig9} shows that they self-organize in the corners of the 2D histogram, i.e., on deterministic behavior, depending on their history of coordinations. It is interesting to note that four groups of frequentists are formed, depending on their coordination with the collective agreements of the learning agents. The widest peak corresponds to the one with the maximum coordination, probably because of the influence of the learning agents. On the other hand, if they are mixed with random walkers, these last ones eventually cluster in the surroundings (in the probability vector phase space) of the collective agreement made by the learning agents. In addition to the random movement from the interactions between them, they are driven by the learning agents to their positions, as they prefer coordination to uncoordination.  

\begin{figure}[ht]
  \includegraphics[width=.48\linewidth]{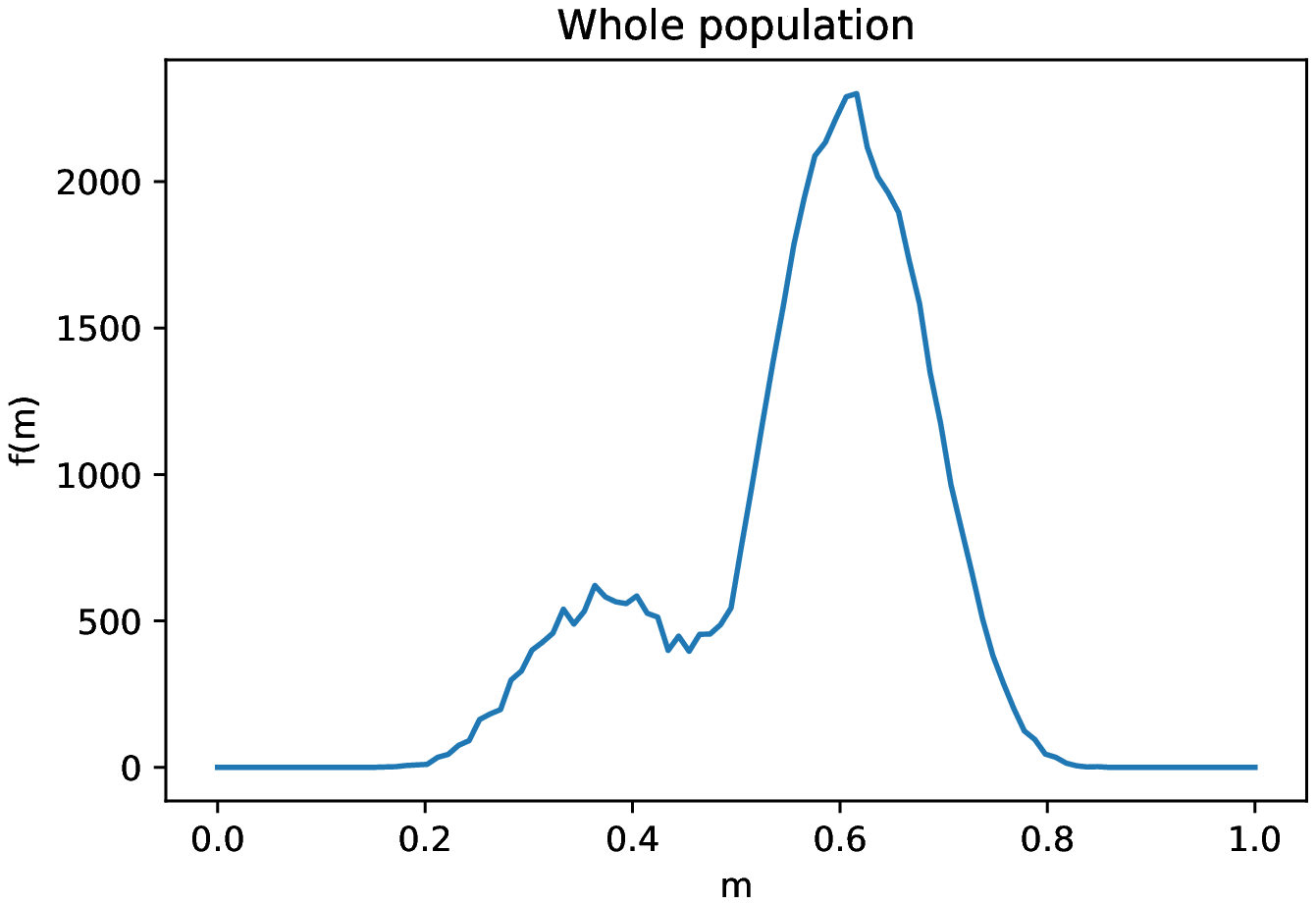}  
  \includegraphics[width=.48\linewidth]{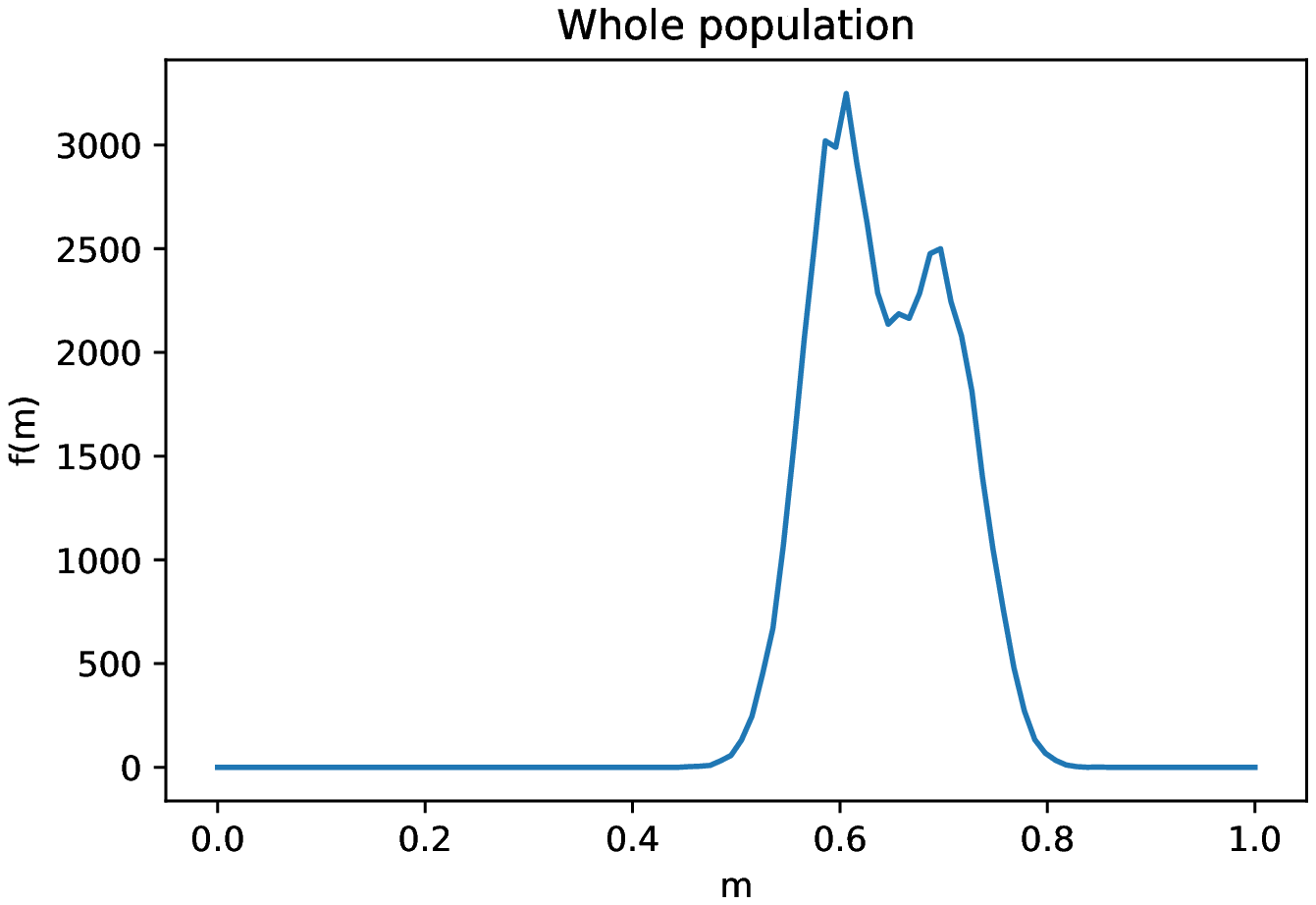}  
  \caption{Coordination ratio for a a frequentist/learning (left) or a random walker/learning (agent)right) population, with the ratio (1:1).}
\label{fig:Fig10}
\end{figure}

\begin{figure}[h]
  \includegraphics[width=.48\linewidth]{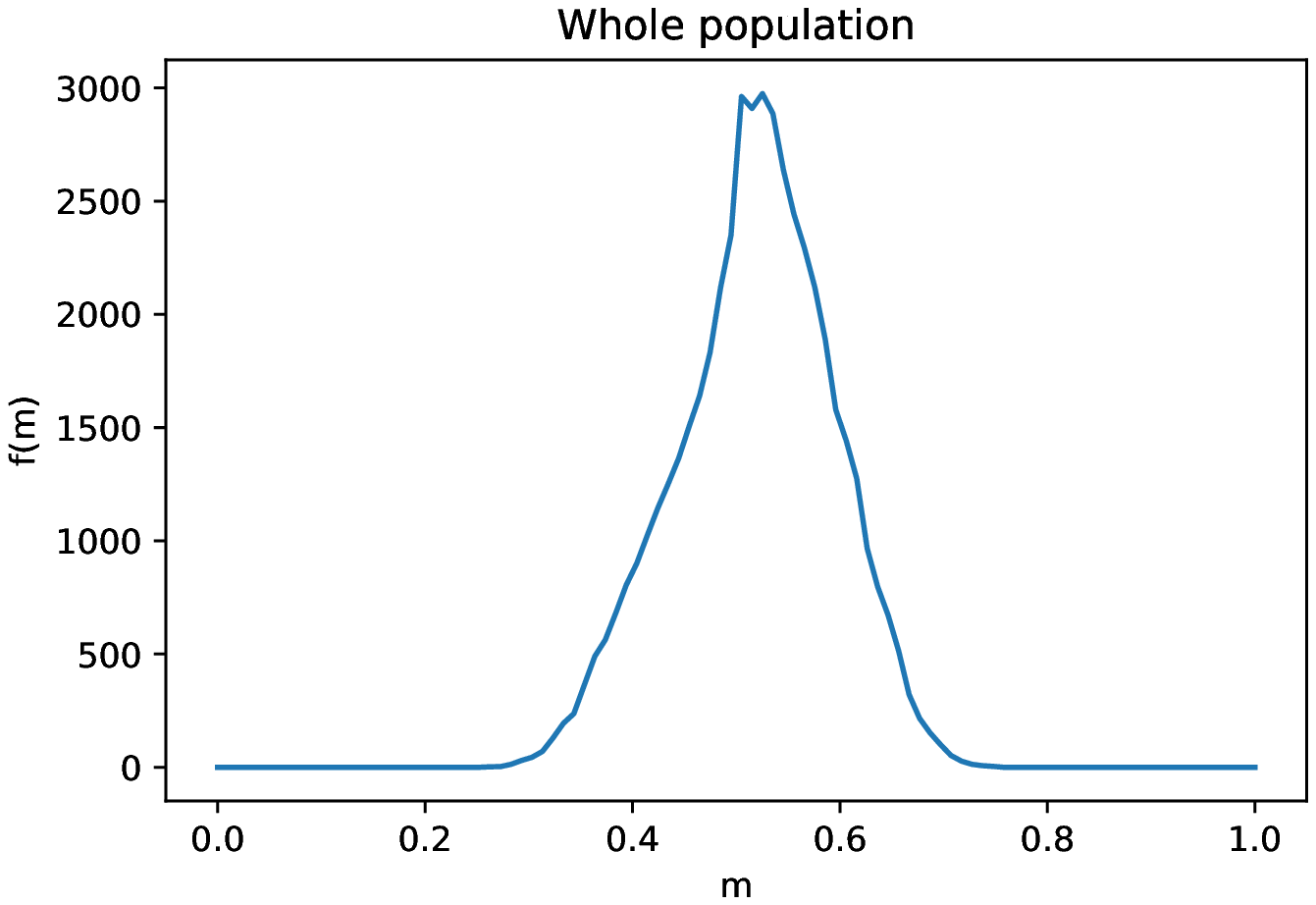}  
  \includegraphics[width=.48\linewidth]{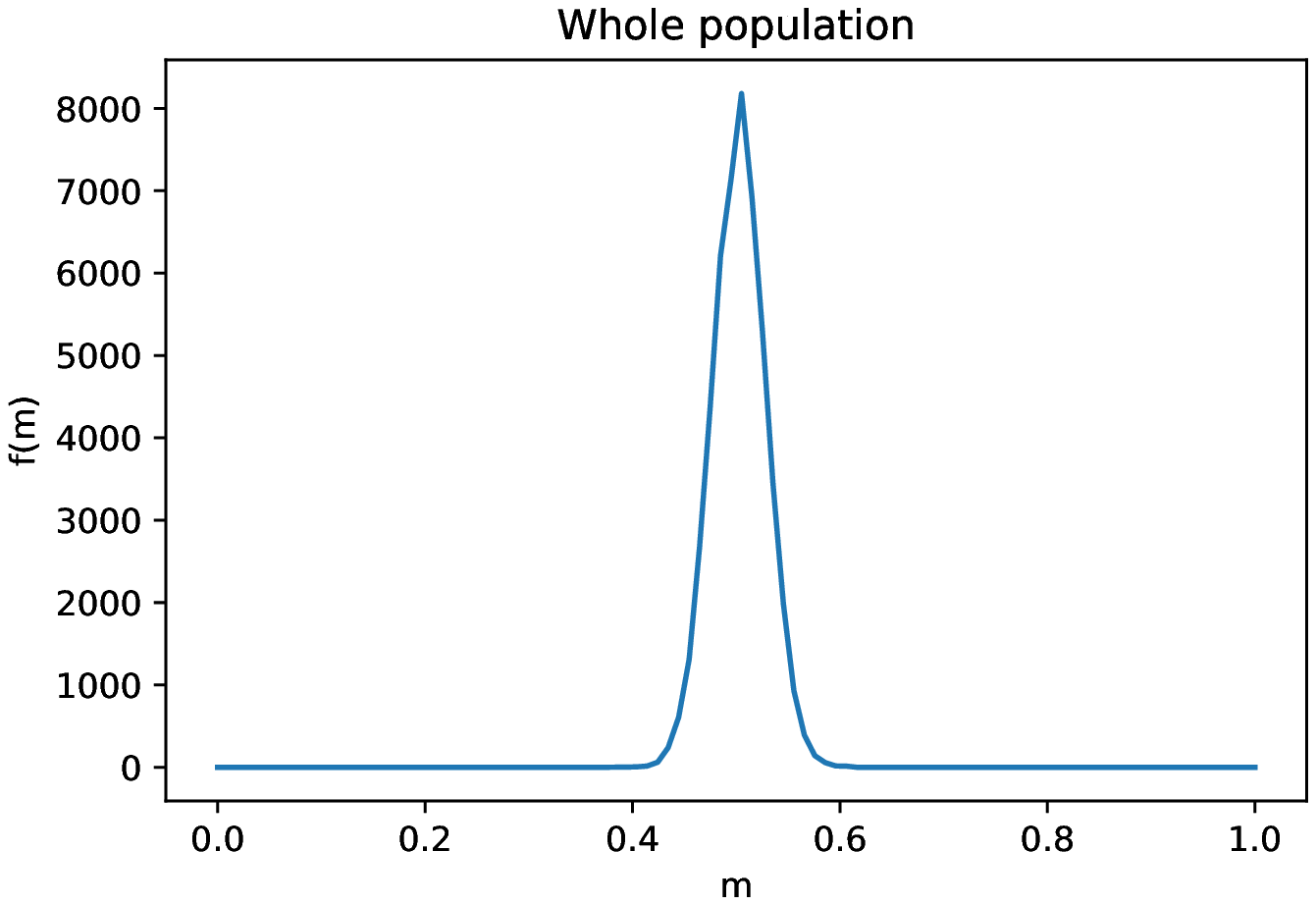}  
  \caption{Coordination ratio for a a frequentist/learning (left) or a random walker/learning (agent)right) population, with the ratio (3:1).}
\label{fig:Fig11}
\end{figure}

Now, in case learning agents are in 1:1 proportion, what we observe in Fig.\ \ref{fig:Fig10} is that 
the inter-marker correlation is destroyed, while the intra-marker still stays, but partially affected too. The coordination ratio is lower than before, but higher than the random value of 0,5. 
If they are mixed with frequentists, they self-organize with an individual criteria that is affected by the presence of learning agents (just like the previous case). The absence of the inter-marker correlation can be noticed in the fact that the coordination ratio for the frequentist population is a sum of two gaussians, one related to the ones that share the learning agents intra-marker correlations and the ones that do not. 
If, on the contrary, they are mixed with random walkers, they stay at the surroundings of the learning agents criteria, as before. However, as the inter-marker correlation has been destroyed, the surroundings are a whole dimension from the probability vector phase space. The coordination ratio is still higher than 0,5 even for random walkers.    

Fig.\ \ref{fig:Fig11} shows results for the situation in which learning agents are a minority ($25\%$). In this case,
both of their correlations are very weak or suppressed. If they are mixed with random walkers, both correlations are destroyed. However, if the majority are frequentists, learning agents stay in the surroundings of the equilibria that frequentists have decided, achieving a slightly better coordination ratio. This occurs because they coordinate with a group of frequentists and have a nonzero probability of coordinating with learning agents and frequentist groups. 

Finally, in a well mixed population of agents with fixed aspirations and three choices, namely $\left\lbrace 0.8,1.1,1.5 \right\rbrace$, the results are similar to the case in which learning agents were in the minority. This means that as before,  they tend to stay in the surroundings of the groups created by frequentists. 
 
\paragraph{Variations on the number of groups and role of migration}

An important question as to the effect and dynamics of behaviors in marked populations, already raised in Ref.\ \citen{mcelreath2003shared}, is the possibility that there are different populations, possibly separated geographically, a situation that may lead to several combinations of markers and actions, not necessarily agreeing between groups of individuals. Therefore, we are now going to consider this issue, and to that end we will consider that 
agents have an individual label indicating to which group they belong to. Following Ref.\ \citen{mcelreath2003shared}, we introduce two migration parameters, \textbf{$m$} and \textbf{$\beta$}, representing the proportion of population that migrates in each group ($m$) and the frequency of the process($\beta$), respectively.  Specifically, migration events take place every $\beta N$ rounds at the end of the corresponding timestep, and in each migration event $m$ individuals in each group are randomly chosen and also randomly reassigned to other groups keeping the population of each group constant. 

As the simplest way to understand the effects of migration, we will consider that we have two separate population groups, denoted by a  binary label $\left\lbrace 0,1 \right\rbrace$.
We have considered medium-sized populations ($N_{group}=250$), while all the other parameters have been set like in previous discussion. 
In what follows we study two cases: two groups with different initial aspirations, so we can assess the role of the spatial structure in the conformation of marked communities, and a second situation with the same fixed initial aspiration for both groups, studying how the amount and velocity of migration influence reaching or not homogeneity between groups in the collective agreements.

\begin{figure}[ht]
  \includegraphics[width=.48\linewidth]{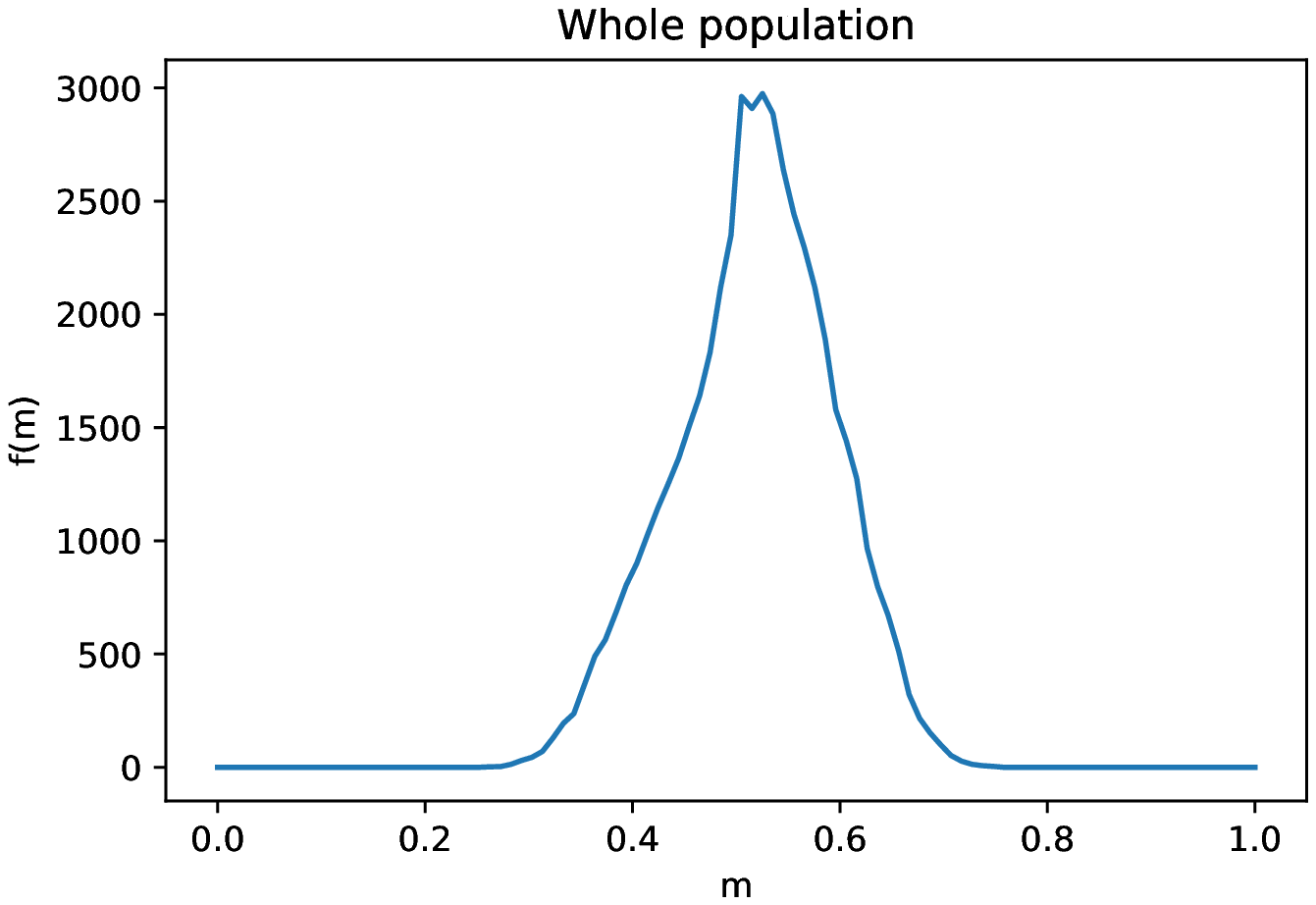}  
  \includegraphics[width=.48\linewidth]{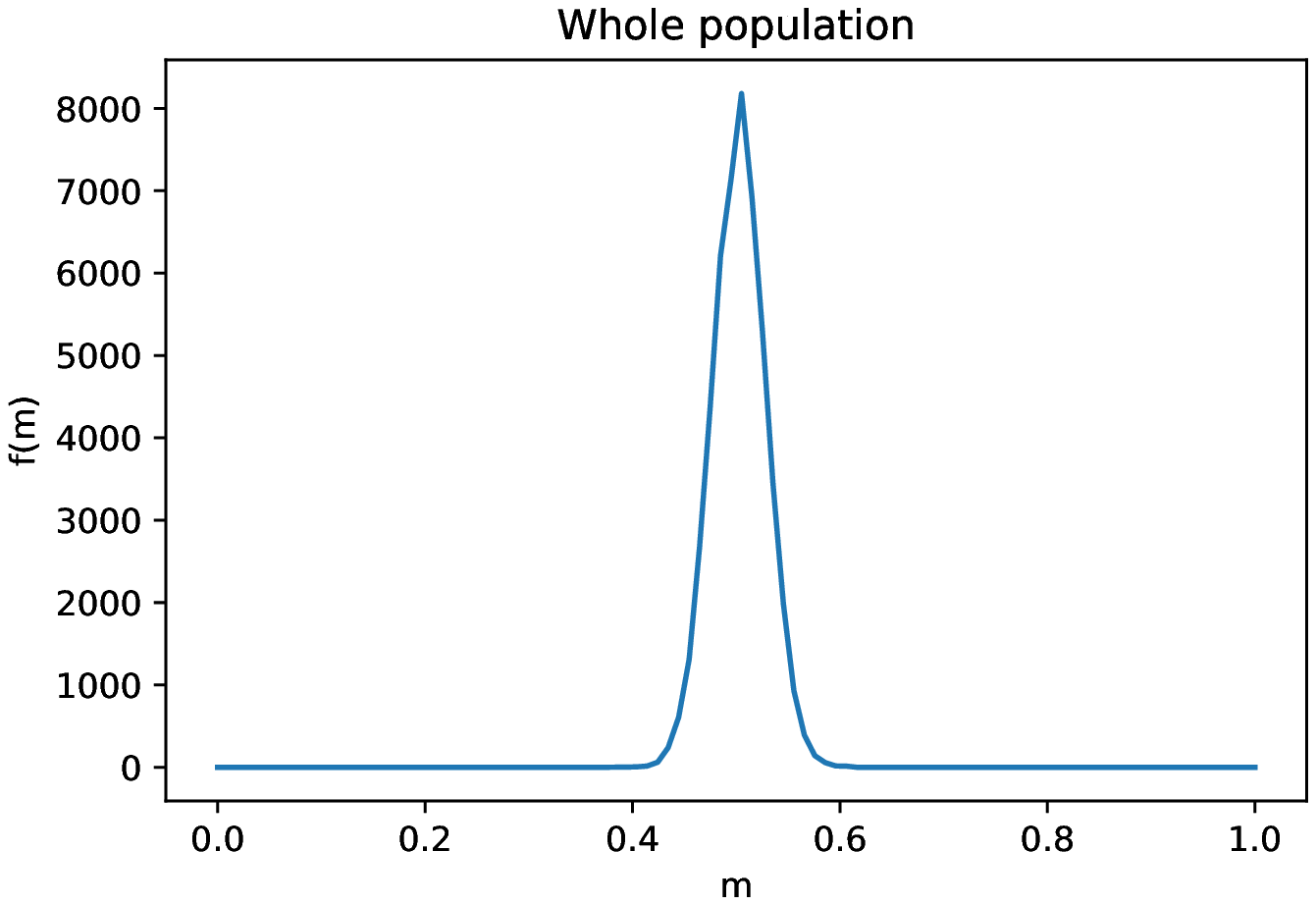}  
  \caption{Coordination ratio for  group 0 (left) and for group 1 (right).}
\label{fig:Fig12}
\end{figure}

For the first study, we have in turn considered two possibilities, one with group 1 with initial aspiration 0.8 and group 2 with initial aspiration 1.1, and another one with group 1 with initial aspiration 1.5 and group 2 with initial aspiration 1.1. Our results are summarized in 
Fig.\ \ref{fig:Fig12}, where it can be seen that the existence of separate groups, which as we said can come from the underlying spatial structure, does not promote new results nor marked communities as, at the end of the day, we obtained two fragmented groups whose main features are basically the same as  the ones studied above. 

For the second study, namely the influence of the migration parameters, we set up  two groups of learning agents with fixed aspirations ($A_{i}=1.1$) with different migration speeds ($\beta$) and a fixed quantity of migrants. To be specific, we will use a proportion of population that migrates in each group $m=1/N$. 
With this setup, we have explored the number of coincidences in the possible criteria for the intra-marker and inter-marker correlations. We define an intra-marker coincidence as the use of the same strategies between groups in relations between agents of the same marker, and inter-marker coincidence as the use of the same strategies between groups in relations between agents of different markers. Therefore, we define the intra/inter-homogeneity as the percentage of realizations for a configuration that promotes these coincidences.

It has to be taken into account that, as we are using $e=0.5$, a $75\%$ of all interactions are intra-marker so it is very likely that this will evolve on a faster time scale. On the other hand, on principle one should expect  that migration leads to more heterogeneity and to longer time scales for the arising of conventions, because agents are exchanged between groups with possibly different actions for the same marker. 
However, our simulations show that this is not the case and migration turns out to be compatible with homogeneity and a single time scale, which means that is not affected by the bias towards interacting with one's own marker. The difference between the numbers for intra and inter homogeneity arises from the fact that the inter-marker correlation is collective agreement in a group, while the intra-marker correlation exists for every marked group (it is proportional to the number of groups). As can be seen in Table \ref{tab:table1}, the ratio 2:1 for these quantities remains more or less constant until the parameters allow to reach  global homogeneity.
\begin{table}[h!]
  \begin{center}
      \begin{tabular}{l|c|r} 
      \textbf{$\beta$} & \textbf{$\%$intra-homogeneity} & \textbf{$\%$inter-homogeneity}\\
      \hline
      no migration & 20 & 45\\      
      10 & 21 & 48\\
      1 & 28 & 60\\
      0.1 & 50 & 100\\
      0.01 & 100 & 100\\
    \end{tabular}
  \end{center}
  \caption{Number of coincidences in collective agreements between different groups for different values of migration speed ($\beta$).}
  \label{tab:table1}
\end{table}
In the former set up we were working with $m=1/N$, just because it is the scenario where differences can arise easily before reaching homogeneity. As expected, if we introduce larger fractions of migrants (larger $m$), larger homogeneity arises. For instance, when $\beta=0.1$, for different $m$ we obtain the results collected in Table \ref{tab:table2}, showing the efficiency of migration to induce homogeneity across groups. 
\begin{table}[h!]
  \begin{center}
    \begin{tabular}{l|c|r} 
      \textbf{$m$} & \textbf{$\%$intra-homogeneity} & \textbf{$\%$inter-homogeneity}\\
      \hline
      no migration & 20 & 45\\      
      $1/N$ & 50 & 100\\
      $2/N$ & 85 & 100\\
      $5/N$ & 100 & 100\\
      $10/N$ & 100 & 100\\
    \end{tabular}
  \end{center}
  \caption{Number of coincidences in collective agreements between different groups for different values of migrating fraction of population ($m$).}
    \label{tab:table2}
\end{table}

\section*{Discussion}

In this work, we have introduced a model for the emergence of marker-related behavior in coordination problems, based on reinforcement learning dynamics, showing that it can provide a framework in order to solve such multi-agent coordination games. Within a certain range of aspirations, it was proved in Ref.\ \citen{macy2002learning} that agents interacting pair-wise develop a collective agreement that promotes a high coordination ratio. We have extended this result here to a population setup showing that, again for some interval of aspirations, the coordination problem is solved and individuals have behavioral rules based on the markers. As was expected from previous works, the learning process leading to correlation between behavior and markers only exists in the region where positive stimuli are statistically stronger than negative stimuli. Therefore, aspirations play a key role in the process of finding an equilibrium. Interestingly, although reinforcement learning is a dynamics that works with local information, correlations are a collective behavior, and in order to reach a globally coordinated equilibrium, it is necessary  a majority of agents that fulfills the conditions on aspirations. 
It is interesting to note also that migration and spatial structure do not promote any new collective effect, and it is only related to homogeneity and isolation effects, depending on the quantity of migrants. While we have considered a single size for our simulations, we have also checked that the results are mostly insensitive to the group size (cf.\ Appendix). This includes smaller groups that might have been the case in early stages of human evolution or in hunter-gatherer societies, showing that our mechanism for the emergence of marker related behavior could also take place there. 

Observable social traits organized in normalized categories are an instance of the concept of several sovereignity centers. In our problem, categories ($=,\neq$) and binary markers ($0,1$) organize two different centers of decision. One around the "equal" category, that defines the intra-marker correlation and one around the "not equal" category, that defines the inter-marker correlation. 
Without markers and categories, the whole population will arrive to an equilibrium with an equal social norm (or coordination strategy) for all. This is what we mean by sovereignty, as observable social traits act now as a new label, that let agents that share this visible trait form a collective agreement on which strategy must be played.

Our model is also informative about the conditions for marker-related behavior to appear. In this respect, it is interesting to note that it is important that aspirations do not change during the process or change very slowly: if habituation\cite{macy2002learning} is introduced, allowing aspiration levels to change as a function of the received payoffs, a large fraction of agents end up with aspirations beyond the range where learning takes place and random behavior arises (cf.\ Appendix), thus breaking the collective behavior related to markers. 
Another important point is the specificity of the game we have studied: in case agents face an asymmetric coordination problem with a preferred (or Pareto-dominant) equilibrium, the most common outcome is that all the individuals end up playing that equilibrium irrespective of the marker, implying that they are then driven by the payoff obtained from the interaction and that the markers become irrelevant. 


It is interesting to consider our work in the light of the pioneering proposal by McElreath {\em et al.}\cite{mcelreath2003shared} Our results are largely different from theirs: Indeed, in our model marker-behavior correlations arise less often, while the existence of more than one group or a spatial structure is not important, and it only leads to more homogeneity between social traits due to migrant exchange between populations. The reason for this is that we are considering a very different dynamics for the actions and, importantly, markers do not evolve in our model. Regarding the actions, we observe that he reinforcement learning dynamics with well tuned aspirations introduces a systematic way for agents to create correlations with markers, while in Ref.\ \citen{mcelreath2003shared} this happens to be an equilibria between different processes of the dynamics. The interpretation of this dynamics is also important; copying the fittest individual's action (and possibly marker) may not be a realistic circumstance, as some social features can not be changed by adaptation, or even if they can, individuals may not have  information about the payoff obtained by every other individual. Reinforcement learning dynamics only uses individual information, avoiding these methodological issues. On the other hand, another difference between the two models is that spatial structure (and then, migration) is not important in order to obtain ethnically marked groups. As we said before, this differentiation is related to the model dynamics, and in our case the only effect that appears joining together different communities is the homogeneizing effect opposed to the possibility of having different conventions when populations are in isolation. \newline
Therefore, it is clear that there is a wide field to explore about the role played by markers in the emergence of group related behavior and the corresponding factors influencing it. Thus, our research paves the way to study new game structures that may represent new social processes, new  marker structure may represent more complex social definitions that the binary one we have used (several features may be added: global markers, continuous ones\dots). Even markers evolving also by reinforcement learning could be incorporated to our setup to study situations in which such change is actually easy in the society. External observable markers are, according to our model, features that may affect the way social norms are described by a society. Agreements are reached inside different groups in the society, and this may in turn lead to a more complex understanding of collective social norms. 


\section*{Acknowledgements}

This research has been funded by the Spanish Ministerio de Ciencia, Innovaci\'on y Uni\-ver\-si\-da\-des-FEDER funds of the European Union support, under project BASIC (PGC2018-098186-B-I00) and by the Comunidad de Madrid, under projects PRACTICO-CM and CAVTIONS-CM-UC3M.

\section*{Author contributions statement}

A. B. and A.S. conceived the research, J. O. wrote the code, conducted the simulation and analyzed the results, and all authors discussed the results and wrote and reviewed the manuscript. 

\section*{Appendix}

\subsection*{Reinforcement learning with habituation}\label{Habituation}

For the sake of completeness, it is interesting to explore the effect of dynamic aspirations. It is quite common that individuals have some aspiration level at the beginning of some interaction process, and that as the interactions proceed they get used to some level of payoff and change their aspirations accordingly. This is generally referred to as habituation, this is, the adaptation of the aspiration to the average payoff obtained by the agent. This is introduced by adding a dynamical rule describing the evolution of the aspiration, given by 
\cite{macy2002learning}:
\begin{equation}\label{eq9}
A_{t+1}=(1-h)A_{t}+h\pi_{t},
\end{equation} 
where $\pi_{t}$ is the payoff obtained by the agent at time $t$. 
The process of updating individual aspirations is introduced in the dynamics for all the timesteps. After a couple of individuals has interacted and collected their payoff, they adapt their aspirations according to the former equation.

In order to understand the possible new effects introduced by this dynamic adaptation of agents, we will consider the case of one initial aspiration for all the population, with values $\left\lbrace 0.8,1.1,1.5 \right\rbrace$ as before, and also two different aspirations in the initial conditions in different proportions. We will use a fast learning approach ($l=0,5$) and  marked-biased interaction ($e=0,5$) in a single group. Our main findings can be summarized as follows. 

When aspirations are initially lower than $1+\delta/2$, and habituation is low (in our case $h<5 \cdot 10^{-3}$), the dynamics modifies slightly the level of aspirations, but it does not change the qualitative results. Correlations exist and they make the agents obtain a high level of coordination, which ends promoting high aspirations (agents reach $A_{eq}\simeq 1+\delta$). 
\begin{figure}[htp]
  \includegraphics[width=.48\linewidth]{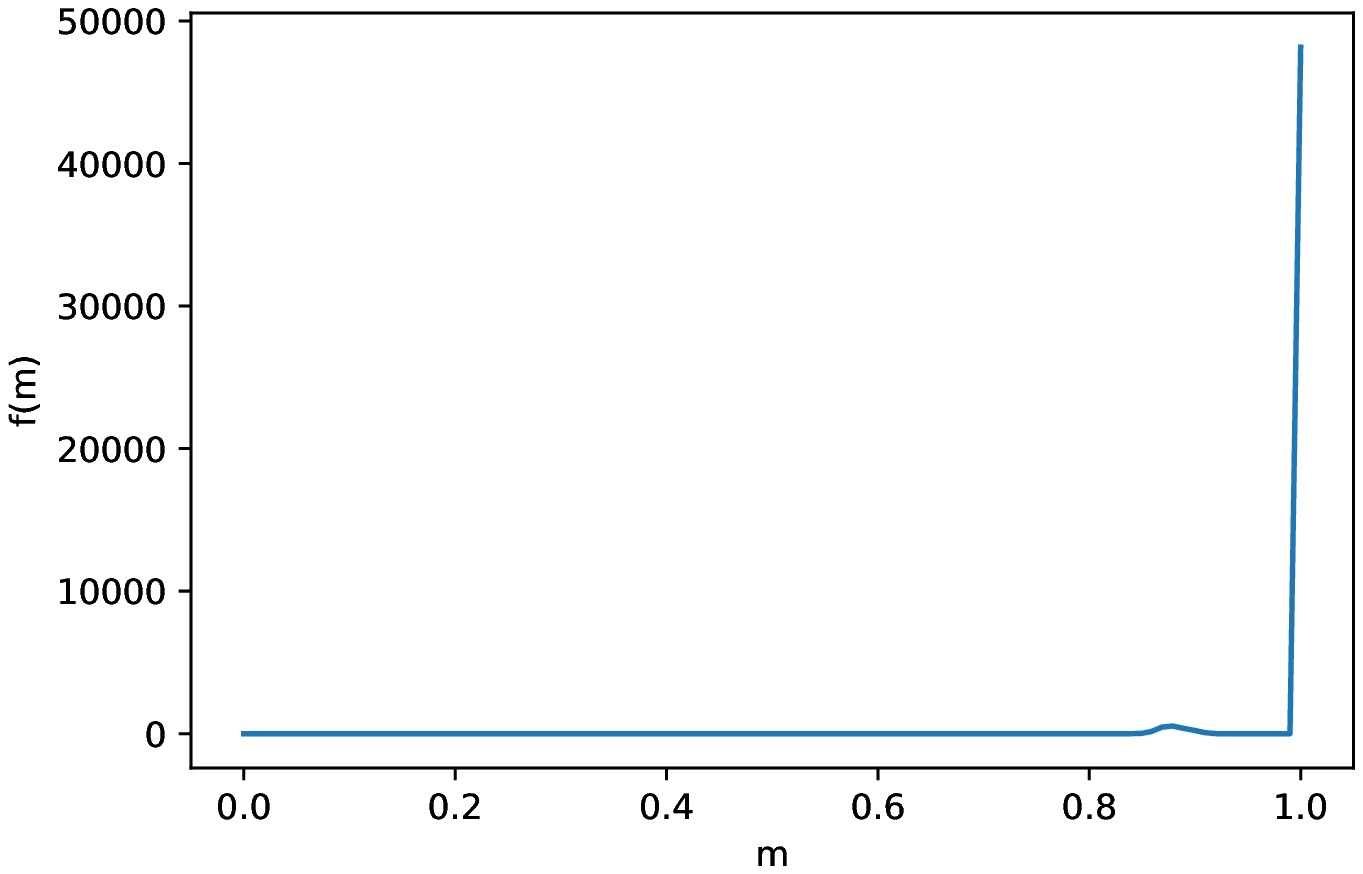}  
  \includegraphics[width=.48\linewidth]{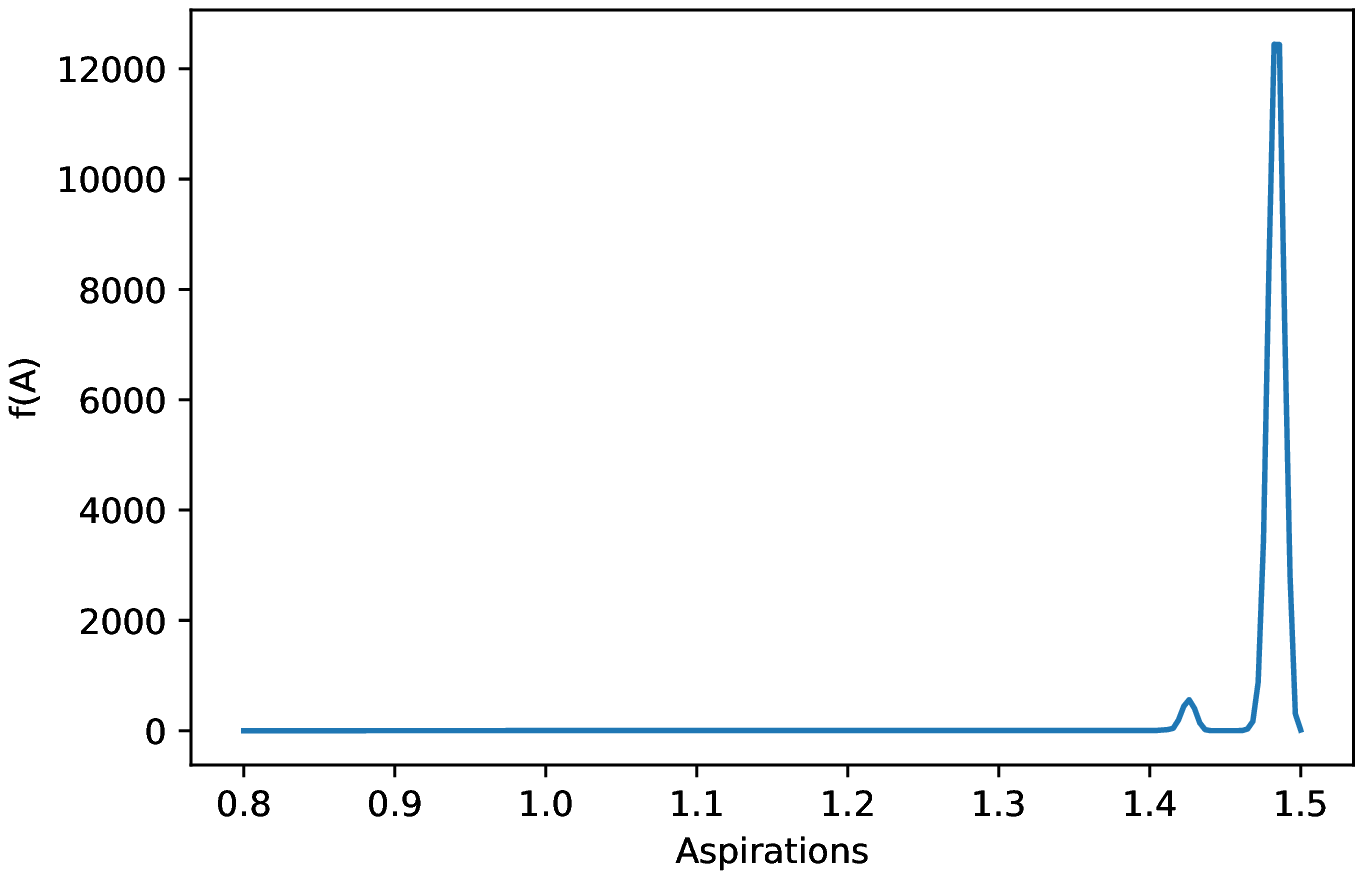}  
  \caption{Coordination ratio  (left) and aspiration (right) distributions. $A_{0}=0,8$ and $h=10^{-3}$}
\label{fig:Fig13}
\end{figure}
Moving now to intermediate values of habituation ($5 \cdot 10^{-3}<h<0,5$), we observe in Fig.\ \ref{fig:Fig14} that habituation destroys correlations, diminishing the coordination ratio and eventually taking the system to uncoordination. As we are biasing the interaction, there is a two-stage destruction. This process happens stochastically, there is not a critical value of $h$ that separates both behaviors. 
Finally, for high values of $h$ (in our case $h\sim 0,5$), it can be seen in Fig.\ \ref{fig:Fig15} the system tends to a uniform distribution of aspirations between the two payoffs, or, in other words, it behaves randomly. 
\begin{figure}[htp]
  \includegraphics[width=.48\linewidth]{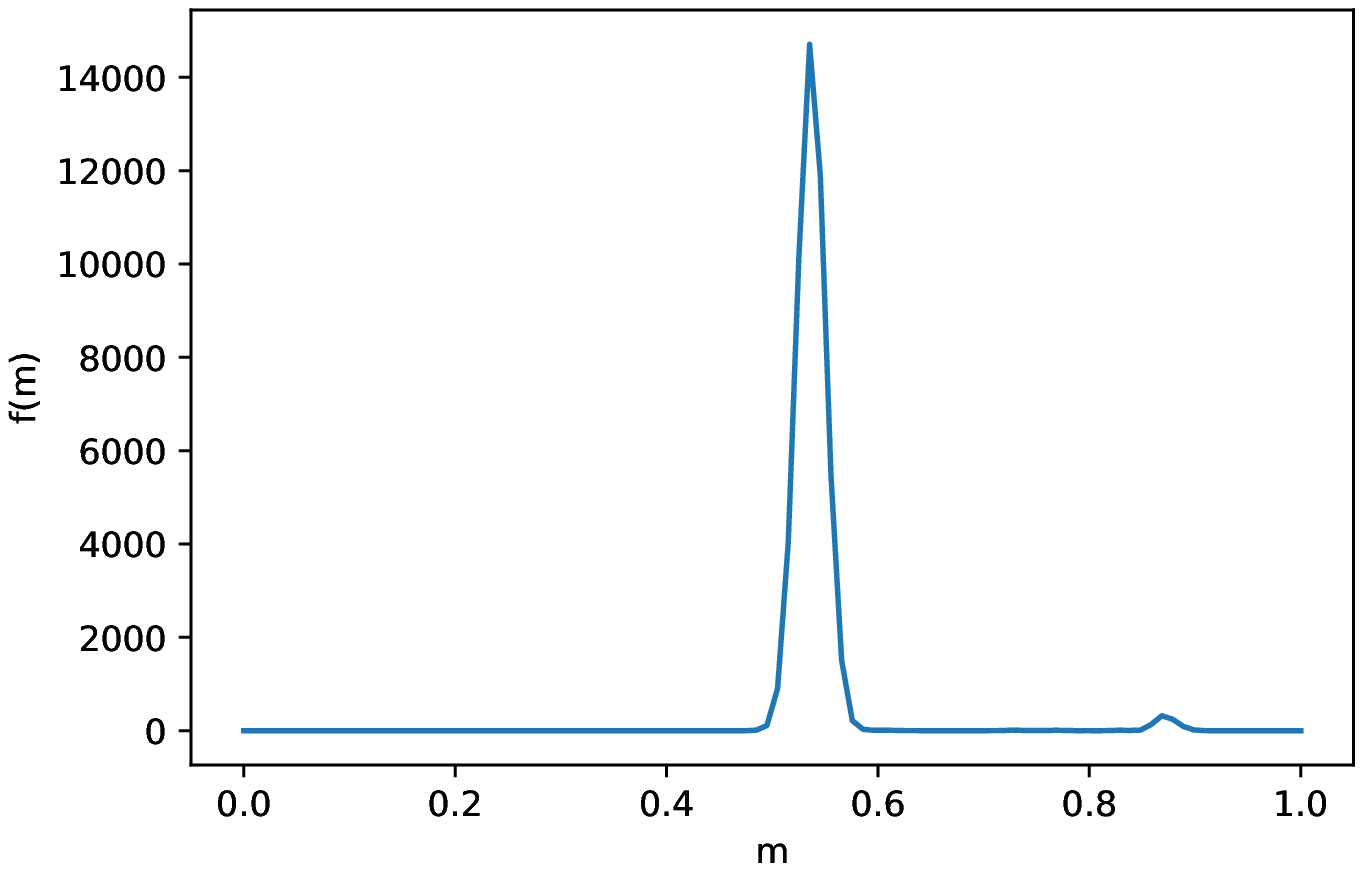}  
  \includegraphics[width=.48\linewidth]{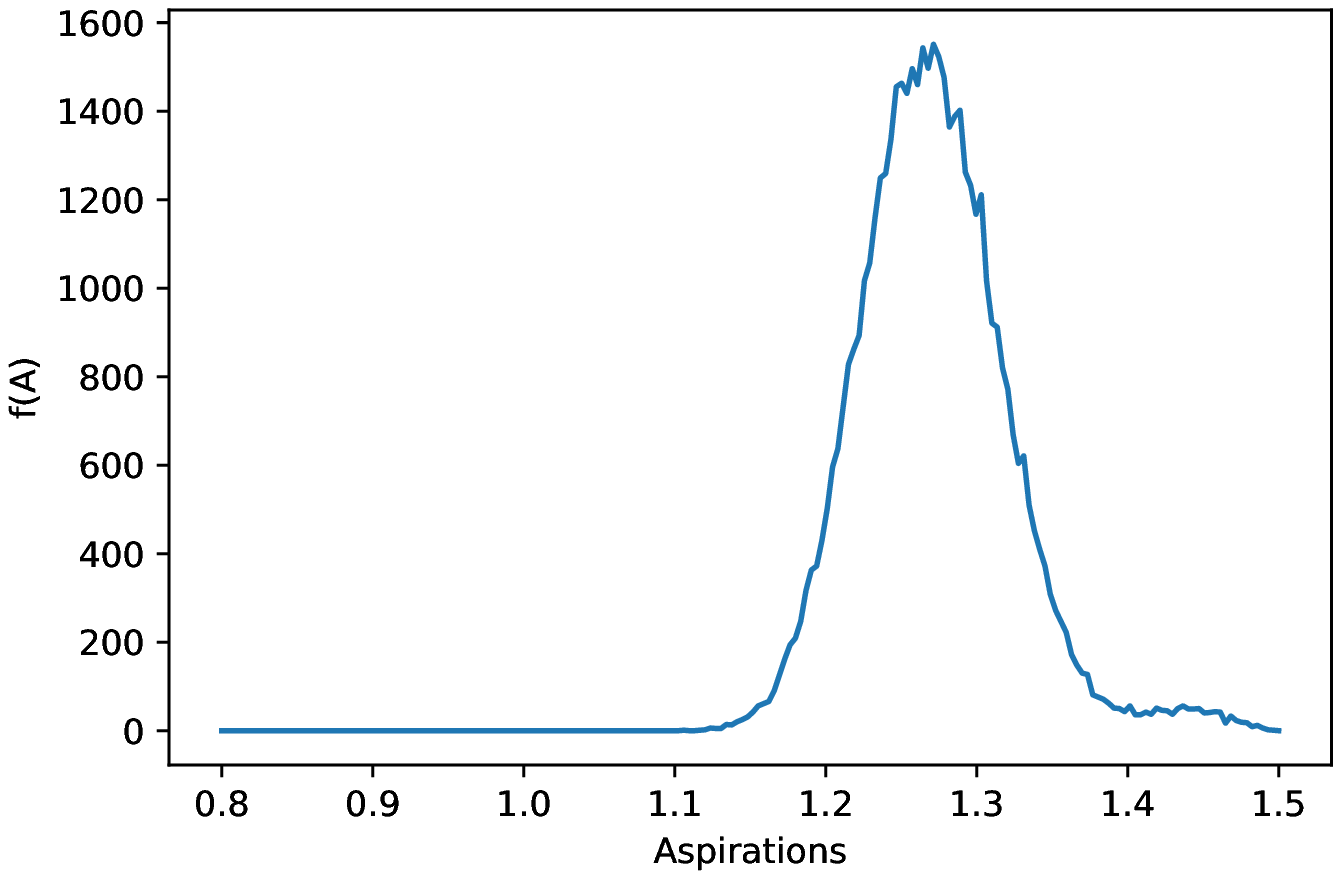}  
\caption{Coordination ratio  (left) and aspiration (right) distributions. $A_{0}=0,8$ and $h=5\cdot 10^{-2}$}
\label{fig:Fig14}
\end{figure}

When aspirations are initially higher than $1+\delta/2$, agents do not create correlations, and their coordination ratio  is about $50\%$, which makes them end in $A_{i}\simeq 1+\delta/2$. All their behavior is randomized, independently of the value of $h$. The only behavior that they share with the previous scenario is the non-markovian one. In addition, the results for fragmented populations are similar to this. As  aspirations are now dynamic, the important fact now is the amount of agents that can create correlations, as this is the behavior that can promote high coordination ratios and high aspirations for the agents. 
\begin{figure}[htp]
  \includegraphics[width=.48\linewidth]{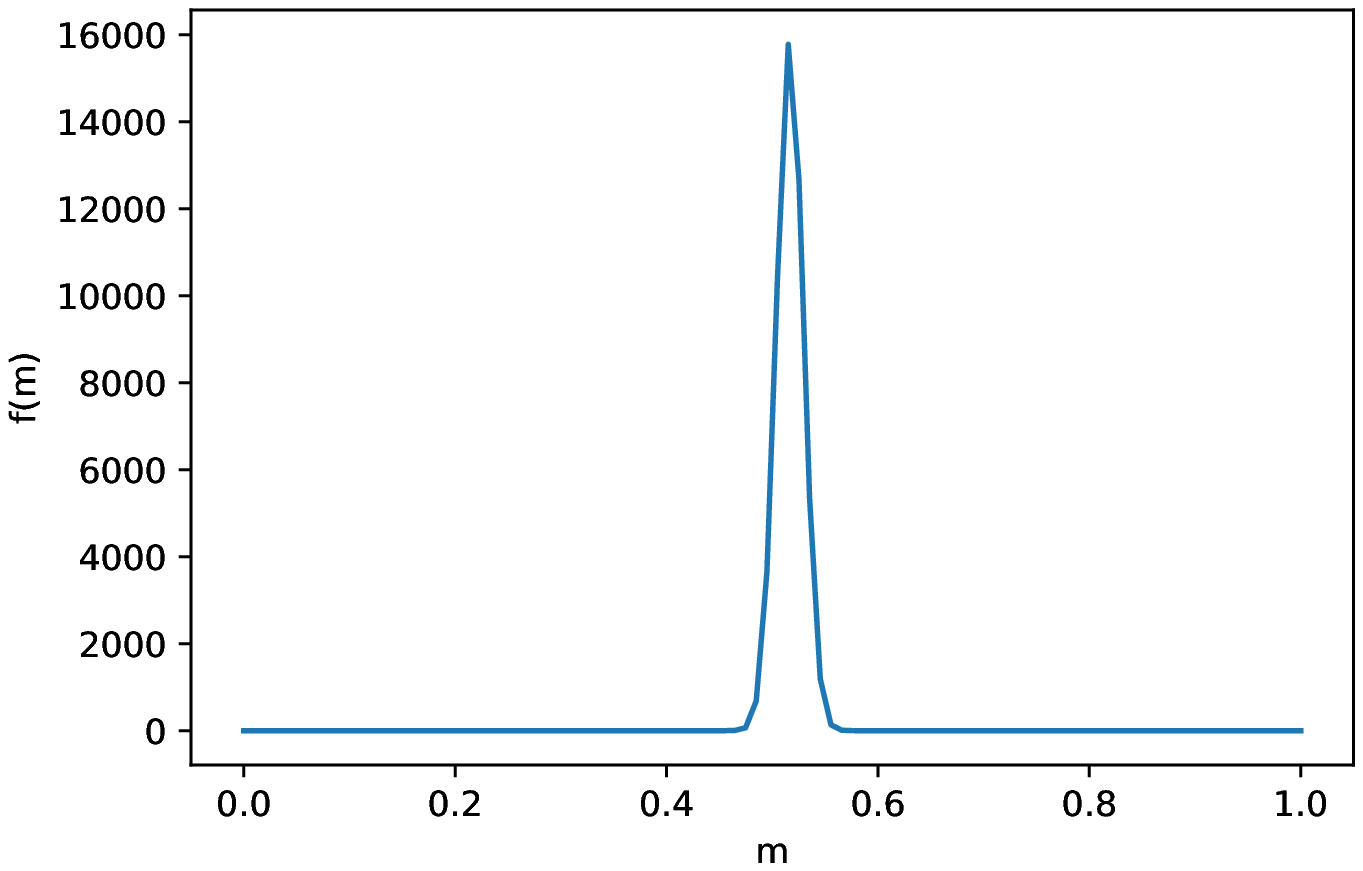}  
  \includegraphics[width=.48\linewidth]{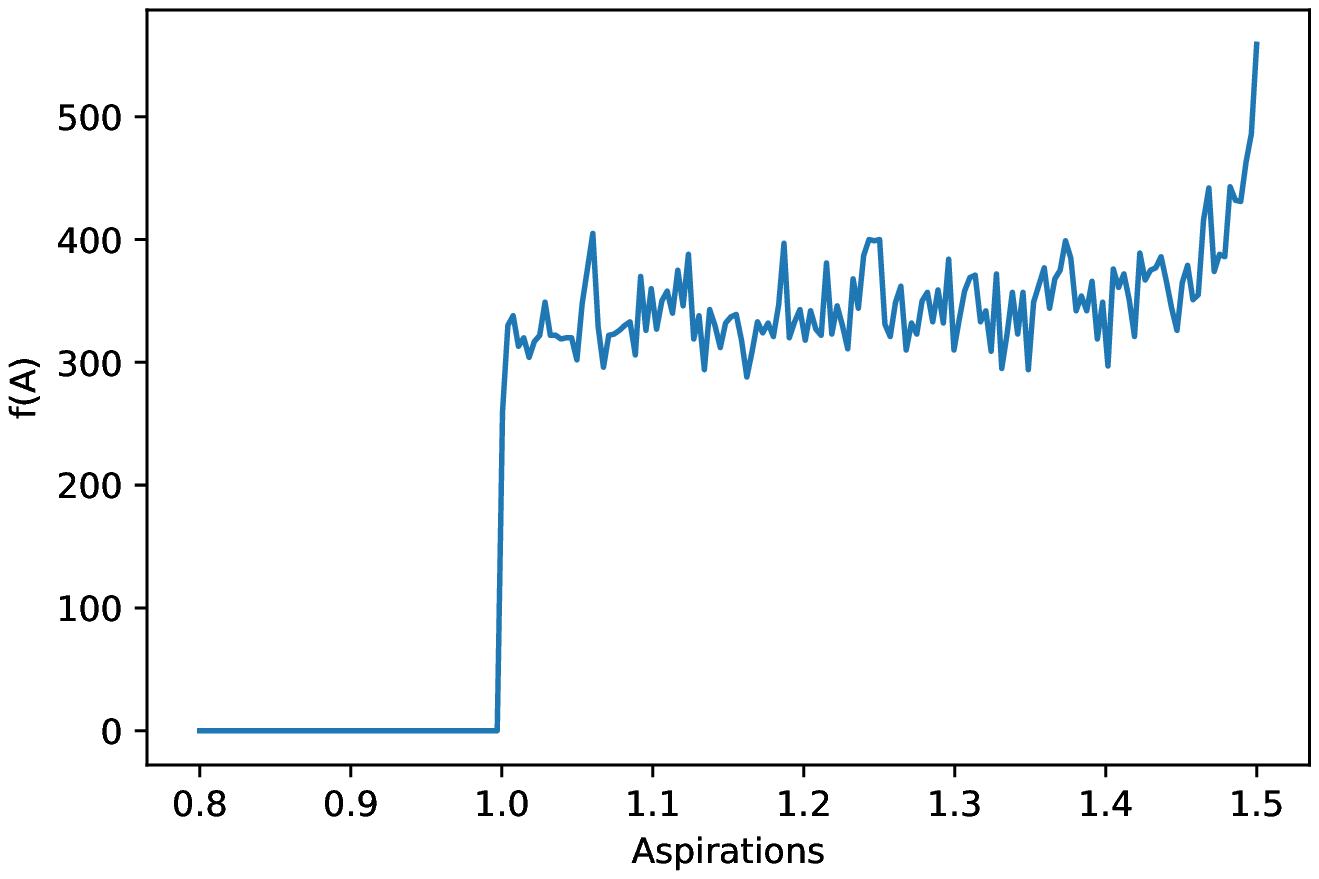}  
\caption{Coordination ratio  (left) and aspiration (right) distributions. $A_{0}=0,8$ and $h=0,5$}
\label{fig:Fig15}
\end{figure}

To sum up, habituation moves aspirations towards the payoff average. This payoff average is $\geq 1+\delta/2$, where correlations can not be created. The specific evolution of the system depends on initial conditions. 

\subsection*{Small groups}

As another feature to complete our study of this model, it is important to consider smaller group sizes for two reasons: First, if the model is to apply to early stages of human history, such as hunter-gatherer groups, the typical number of individuals involved in the process may certainly be smaller than the one considered so far. Second, it is possible to check the model predictions and assess how applicable the model is to actual situations by using 
experimental studies of our setup, and small numbers are easier to handle in that case. After carrying out simulations for smaller groups (Number of agents = $\left\lbrace 200, 500, 1000, 1250, 1500 \right\rbrace$), qualitative results do not change. This includes the effects arising from the variations on the learning rate, biasing, habituation and fragmented aspirations are the same, as they do not affect the stimuli structure. 

\begin{figure}[htp]
\includegraphics[width=.48\linewidth]{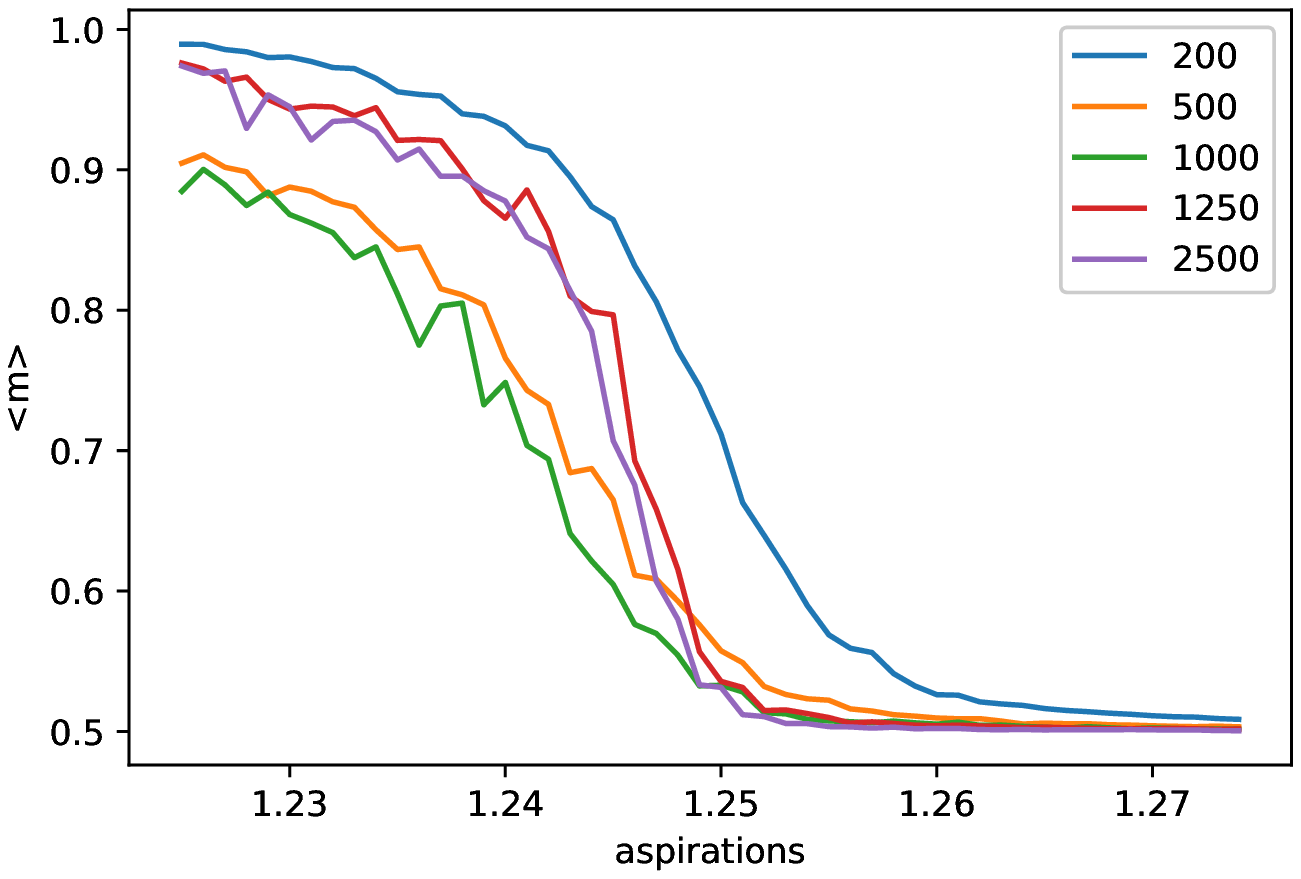}  
\includegraphics[width=.48\linewidth]{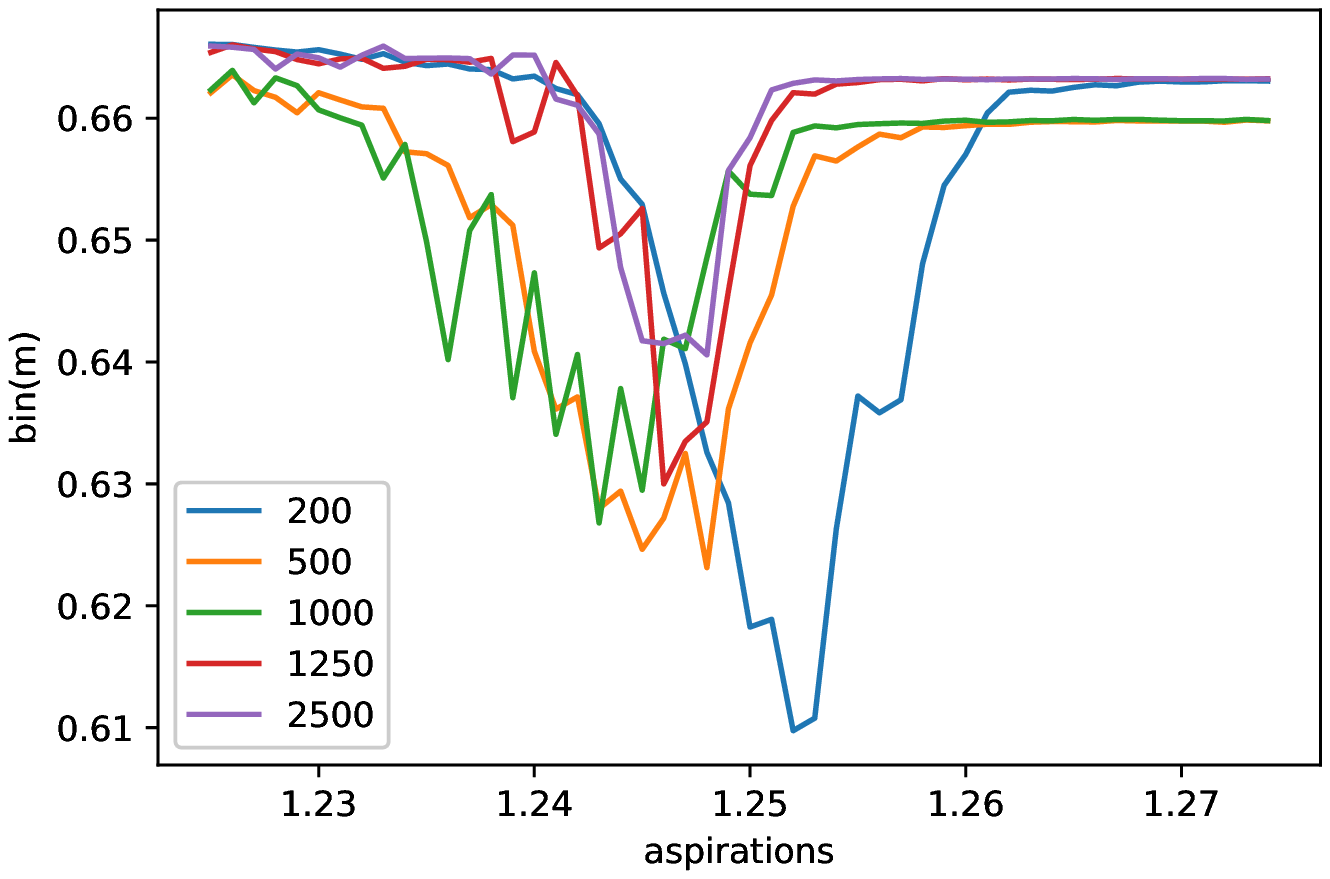}  
\caption{Average coordination ratio (left) and Binder cumulant (right) for several sizes in the $1+\delta/2$ region.}
\label{fig:Fig17}
\end{figure}

However, the smaller size may give rise to smoothening of the transitions between regimes that we observed in Fig.\ \ref{fig:Fig3}, and one could in fact ask if there is some kind of Ising-like transition \cite{PhysRev.65.117} between the different regimes we have described. Fig.\ \ref{fig:Fig17} presents a plot of the region around $A_{i}=1+\delta/2$ for several sizes of the system:
We will focus on the learning-random walker region ($A_{i}\simeq 1+\delta/2$). The other important region ($A_{i}=1$) has an easier analysis. The peak in the variance tends to be smaller as we increase the size, giving signals of a finite size effect. There are more reasons, related with the variation of the learning rate and the behavior of frequentists. The region ($A_{i}\simeq 1+\delta/2$) promotes a qualitative change of behavior, which we could not relate to any parameter of the system. However, as we can see in the picture below, the system does not exhibit the typical finite size scaling one would expect if we were facing a face transition.

\end{document}